\documentclass[10pt,journal,compsoc]{IEEEtran}

\usepackage{fancyhdr}
\usepackage{microtype}
\usepackage{lipsum}
\usepackage{setspace}
\usepackage{flushend}
\usepackage{balance}

\usepackage{amsmath, amssymb, amsthm}
\usepackage{mathtools}
\usepackage{bm}

\usepackage[noend]{algpseudocode}
\usepackage{algorithm}
\usepackage{listings}

\usepackage{graphicx}
\usepackage[table]{xcolor}
\usepackage{colortbl}
\usepackage{booktabs}
\usepackage{multirow}
\usepackage{caption}
\usepackage{subcaption}
\usepackage{varwidth}
\usepackage{siunitx}
\usepackage{url}
\usepackage[hidelinks]{hyperref}

\usepackage{enumitem}
\usepackage{xspace}

\newcommand{\acro}{{\textsc{Blest}}\xspace}

\begin{document}

\title{Graph Traversal on Tensor Cores: A BFS Framework for Modern GPUs}

\author{Deniz~Elbek
        and Kamer~Kaya
\thanks{1) Deniz Elbek (\texttt{deniz.elbek@sabanciuniv.edu}) and Kamer Kaya
(\texttt{kaya@sabanciuniv.edu}) are with the Department of Computer
Science and Engineering, Faculty of Engineering and Natural Sciences,
Sabancı University, Istanbul, Turkey, and also with the Center of Excellence in Data Analytics, Sabancı University, Istanbul, Turkey.\\2) Kamer Kaya is the corresponding author of this work.}
}

\maketitle

\begin{abstract}
Modern GPUs have Tensor Cores~(TCs) capable of extremely high-throughput matrix operations, yet graph algorithms remain difficult to accelerate because of their irregular and data-dependent execution patterns. This work presents \acro, a TC–accelerated framework that reformulates Breadth-First Search (BFS) as a bit-level sparse matrix–vector computation while addressing the load imbalance, memory inefficiency, and synchronization overheads that limit prior approaches. \acro introduces Binarized Virtual Slice Sets (BVSS), a graph representation that partitions work into balanced warp-level units and schedules only frontier-relevant regions of the graph. It further employs an optimized TC layout that maps neighbour checks onto binary MMA instructions without wasted outputs, reducing the number of required MMA calls by 8× compared with prior layouts. To mitigate atomic and cache bottlenecks, \acro{} incorporates a lazy vertex-update scheme. We revisit the switching terminology for BFS and propose a mechanism that dynamically transitions from TCs to CUDA cores when it becomes more efficient. We also extend \acro{} to multi-source BFS and closeness centrality workloads. Finally, we introduce a scalable graph reordering method that improves compression for scale-free-like graphs, while using RCM to improve locality for others. Across a broad set of real-world graphs, \acro achieves average speedups of 22.0$\times$, 7.7$\times$, 8.1$\times$, and 5.9$\times$ over GAP, Gunrock, GSWITCH, and BerryBees, respectively, establishing a new BFS baseline on GPUs. Thanks to its high performance, \acro can compute the exact closeness centralities of 65.6M vertices in a social network with 3.6B edges in an hour using 100 H100 GPUs.

\end{abstract}


\IEEEpeerreviewmaketitle

\section{Introduction}
\label{sec:intro}

\IEEEPARstart{M}{odern} GPUs have specialized matrix-multiply-accumulate (MMA) units, i.e., Tensor Cores~(TCs) that deliver high throughput for matrix operations. These units are designed for dense and highly regular workloads. Graph algorithms, however, live at the opposite end of the spectrum: they are driven by sparse, irregular, and data-dependent access patterns, and their performance is constrained not by peak arithmetic throughput but by memory behavior, synchronization, and load imbalance. Therefore, for graphs, the main question is not whether MMA units are fast, but under what conditions their dense computational model can be exploited despite the irregularity inherent in state-of-the-art algorithms. In this work, our main focus will be on Breadth-First Search (BFS), however, we will also briefly discuss the use of TCs for other (iterative) graph algorithms.

As one of the most important algorithms on graphs, BFS has
applications across a broad range of domains---including
network science~\cite{Takes2011}, computer networks~\cite{Akram2013}, recommendation systems~\cite{Pandey2017},
and compiler design~\cite{Barnat2003}---where the BFS performance has a direct, end-to-end impact on the overall
application throughput. Therefore, it is a popular research area in high-performance computing~(HPC), and its performance has been optimized both
algorithmically~\cite{Leiserson2010,Direction,Dhulipala2021} and
implementation-wise across many systems, including
CPUs~\cite{Direction,Shun2013}, GPUs~\cite{Merrill2015,gunrock,gswitch}, and clusters~\cite{Buluc2011,Checconi2014}.

A single BFS can be represented as an iterative
sparse matrix--sparse vector multiplication
(\mbox{{\tt SpMSpV}})~\cite{GraphBLAS}, where the matrix
corresponds to the transpose of the bit-adjacency matrix of the underlying graph, and the vector is a bit-frontier vector with set bits identifying the nodes in the current frontier. Each  {\tt SpMSpV} produces an output frontier (bit) vector for the next level. The process continues until the frontier is all-zero, i.e., no reachable unvisited node remains. To harness the TCs, the multiplication must be recast as a specialized form of GEMM. Recent work~\cite{berrybees} has introduced a methodology that successfully integrates the TCs into BFS pipelines and achieves speedups over the state-of-the-art across a diverse set of graphs. Nevertheless, several inefficiencies and extensions remain, which we address in this work. Beyond proposing the fastest BFS framework to date, we systematically analyze using TCs for graph algorithms. Our contributions are 5-fold:

\begin{enumerate}[leftmargin=*, label=\textbf{\arabic*.}]
  \item We propose \acro, which, to the best of our knowledge, is currently the fastest BFS implementation in the literature. 
  \begin{itemize} \item Thanks to its novel structure, {\em Binarized Virtual Slice Sets}~(BVSS), it has {\em{near-perfect}} inter-warp load balance.
  \item \acro employs a TC multiplication layout that {\em{optimally}} maps the required edge operations onto these dense units, reducing the number of required MMA calls
  by a factor of $8\times$ compared to state-of-the-art. 
  \item It introduces a lazy update scheme that (i) reduces
  the inherent cost of atomics and (ii) significantly improves
  cache locality across all levels of the memory hierarchy.
  \item \acro fuses the per-level BFS kernels into a single persistent
  kernel to eliminate host-side synchronization and kernel-launch
  overhead.
  \end{itemize}

  \item We reconsider {\em switching} in the TC era to specialize each BFS level execution, and enhance \acro to perform a transition between TCs and CUDA cores when necessary.

  \item We extend \acro to multi-source BFS workloads, altering BVSS
  to support that transition as
  efficiently as possible. We evaluate these workloads on a
  representative multi-source BFS application: {\em closeness centrality}.


  \item We introduce \textsc{JaccardWithWindows}, a novel yet simple reordering algorithm applicable to the largest real-world graphs. We use it for scale-free-like graphs to increase the effective compression ratio of BVSS. For other graphs, we use a traditional reordering algorithm, Reverse Cuthill--McKee~(RCM)~\cite{rcm}, to reduce L1 cache miss rates and excessive L2 cache line fetches. 

  \item We evaluate single-/multi-source \acro on a broad set of real-world graphs and
  achieve average speedups of $21.96\times$, $7.70\times$, $8.12\times$, and $5.93\times$ over GAP~\cite{gap}, Gunrock~\cite{gunrock}, GSWITCH~\cite{gswitch}, and BerryBees~\cite{berrybees}, respectively.

\end{enumerate}

The article is organized as follows. Section~\ref{sec:background} provides background on BFS, TCs, and introduces the notation used throughout the paper.  Section~\ref{sec:ds} introduces BVSS, our novel data structure for BFS on TCs. Section~\ref{sec:ordering} presents the two reordering strategies employed in this work. Section~\ref{sec:compute} describes \acro's compute pipeline for single-source BFS and Section~\ref{sec:extension} extends it to multi-source BFS workloads. Both pipelines are evaluated extensively in Section~\ref{sec:experiments}. Related work is discussed in Section~\ref{sec:related}, and Section~\ref{sec:conclusion} concludes the article.

\section{Background and Notation}
\label{sec:background}

Let $G = (\mathcal{V}, \mathcal{E})$ be a directed graph with $n = |\mathcal{V}|$ vertices and $m = |\mathcal{E}|$ edges, and let $\mathbf{A} \in \{0,1\}^{n \times n}$ denote the transposed adjacency matrix, so that $\mathbf{A}[i][j] = 1$ if and only if $(j, i) \in \mathcal{E}$. Let $\mathbf{x}^{(k)} \in \{0,1\}^{n}$ be the bit-frontier vector at BFS level $k$, where $x^{(k)}_i = 1$ if vertex $i$ belongs to the frontier at level $k$. Let
$
\mathbf{v}^{(k)} = \bigvee_{\ell=0}^{k} \mathbf{x}^{(\ell)}
$
denote the visited vector after level \(k\). The per-level BFS computation advances the frontier via
\begin{equation}
\begin{aligned}
    \mathbf{x}^{(k+1)}
    &=
    \left(\mathbf{A} \cdot \mathbf{x}^{(k)}\right)
    \land \neg \mathbf{v}^{(k)}, \\
    &=
    \left[
    \left(
    \bigvee_{j=0}^{n-1}
      \mathbf{A}[i,j] \land x^{(k)}_j
    \right)
    \land \neg v^{(k)}_i
    \right]_{i=0}^{n-1},
\end{aligned}
\label{eq:spmspv}
\end{equation}
where the multiplication operates over the Boolean semiring $(\vee,\, \wedge)$; i.e., the dot product over $(+, \times)$ is replaced by $(\vee, \wedge)$, and already visited vertices are filtered from the result. The expression evaluates to $1$ if and only if there exists at least one incoming neighbour $j$ of $i$ such that $\mathbf{A}[i][j]=1$, $x^{(k)}_j = 1$, and $i$ has not been visited before, i.e., $x^{(\ell)}_i = 0$ for all $\ell \leq k$. For a source $s \in \mathcal{V}$, the recurrence in \eqref{eq:spmspv} is initialized with $x^{(0)}_s = 1$, and $x^{(0)}_i = 0$, $\forall i \neq s$, and terminates at the first level $k^{*}$ for which $\mathbf{x}^{(k^{*})} = \mathbf{0}$. The algorithm returns an array $\mathtt{level} : \mathcal{V} \to \mathbb{Z}_{\geq 0}$, where $\mathtt{level}[i] = k$ such that $x^{(k)}_i = 1$. For an unreachable $i$, $\mathtt{level}[i] = \infty$, so $x^{(k)}_i = 0$ for all $k$.

\subsection{GPU Tensor Core Architecture}
\label{sec:arch}

GPUs comprise multiple compute units called Streaming Multiprocessors (SMs), each subdivided into SM Sub-Partitions (SMSPs). On recent NVIDIA GPUs, each SMSP contains one TC, and a warp---a team of 32 threads executing the same instruction under the Single Instruction Multiple Thread (SIMT) paradigm---scheduled onto that partition can dispatch its work onto the partition's private TC. 

TCs support multiple precisions and even structured sparsity configurations; however, their sparsity support imposes strict structural requirements not suitable for the irregular sparsity patterns of graphs. Still, TCs have found their way into a variety of domains that they were not originally designed for, including SpMV~\cite{bitmap_spmv, dasp_spmv}, SpMM~\cite{dtc_spmm, torsten_spmm, acc_spmm, brp_spmm, cute_spmm, flashsparse_spmm, groot_spmm, fastSpmm, voltrix, jigsaw, hr_spmm, nm_spmm, hcspmm, rshspmm, mpspmm}, triangle counting~\cite{triangle_counting}, quantized SpMM~\cite{quantized}, fully-homomorphic encryption~\cite{neo}, epistatic detection~\cite{epiclear}, graph neural networks~\cite{tc_gnn, gnn_sptc}, sparse deep learning~\cite{deepneural_spmm, venom, sparsetir, sparta}, sparse LLM inference~\cite{coruscant, generalsparse, flashllm, spinfer}, stencil~\cite{stencil}, reduction~\cite{reduction}, and more recently BFS~\cite{berrybees}. 

TCs can operate on the bit-level: given two $k$-bit vectors, a TC  computes their product over the $(\mathtt{popc}, \wedge)$ semiring, where $\mathtt{popc}$ returns the number of set bits produced by the bitwise \textsc{and} of its two operands. Although Eq.~\eqref{eq:spmspv} requires the Boolean semiring $(\vee, \wedge)$, it can be faithfully emulated via the $(\mathtt{popc}, \wedge)$ semiring supported by the hardware. Specifically, let $\mathbf{a}_i \in \{0,1\}^{n}$ denote the $i$-th row of $\mathbf{A}$; then
\begin{equation}
    x^{(k+1)}_i = \mathbf{1}\!\left[\mathtt{popc}\!\left(\mathbf{a}_i \wedge \mathbf{x}^{(k)}\right) > 0\right].
    \label{eq:popc}
\end{equation}
The emulation is exact: $\mathtt{popc}(\mathbf{a}_i \wedge \mathbf{x}^{(k)}) > 0$ if and only if $\bigvee_{j}\left(\mathbf{a}_i[j] \wedge x^{(k)}_j\right) = 1$. Here, the operands are not plain $k$-bit vectors, but $m \times k$ and $k \times n$ matrices, whose dimensions depend on the chosen precision configuration. Three binary MMA shapes are supported on recent NVIDIA architectures: \texttt{m8n8k128}, \texttt{m16n8k128}, and \texttt{m16n8k256}~\cite{ptx}. In this work, we utilize the smallest available configuration, \texttt{m8n8k128}, to achieve the finest computation granularity.

\subsection{Directions, Updates, and Switching in BFS}
\label{subsec:terminology}

BFS is driven entirely by {\em information transmission}: the {\em visited} state of the vertices in the current frontier is passed to next frontier vertices. Following the terminology of Beamer~et~al.~\cite{Direction}, in a level-synchronized BFS, this transmission can happen in two directions: (1) In {\bf{\em top-down}} exploration, the vertices in the current frontier transfer their visited states {\em to} their unvisited (outgoing) neighbours. (2) On the other hand, in the {\bf{\em bottom-up}} exploration, the unvisited vertices transfer the visited states {\em from} their (incoming) neighbour vertices in the current frontier. Hence, in top-down, the work queue is the current frontier, and in bottom-up, it is the set of unvisited vertices. The literature calls the former exploration direction {\bf{\em push-based}} since the information is {\em pushed} from the current frontier to the next one. Similarly, the latter exploration is called {\bf{\em pull-based}}. This coupling sounds natural since, in a top-down BFS, a thread naturally pushes updates to a shared state, whereas in a bottom-up one, it pulls the (incoming) neighbours' states to its private state.

\begin{figure*}[!htbp]
     \centering
     \includegraphics[width=0.97\textwidth]{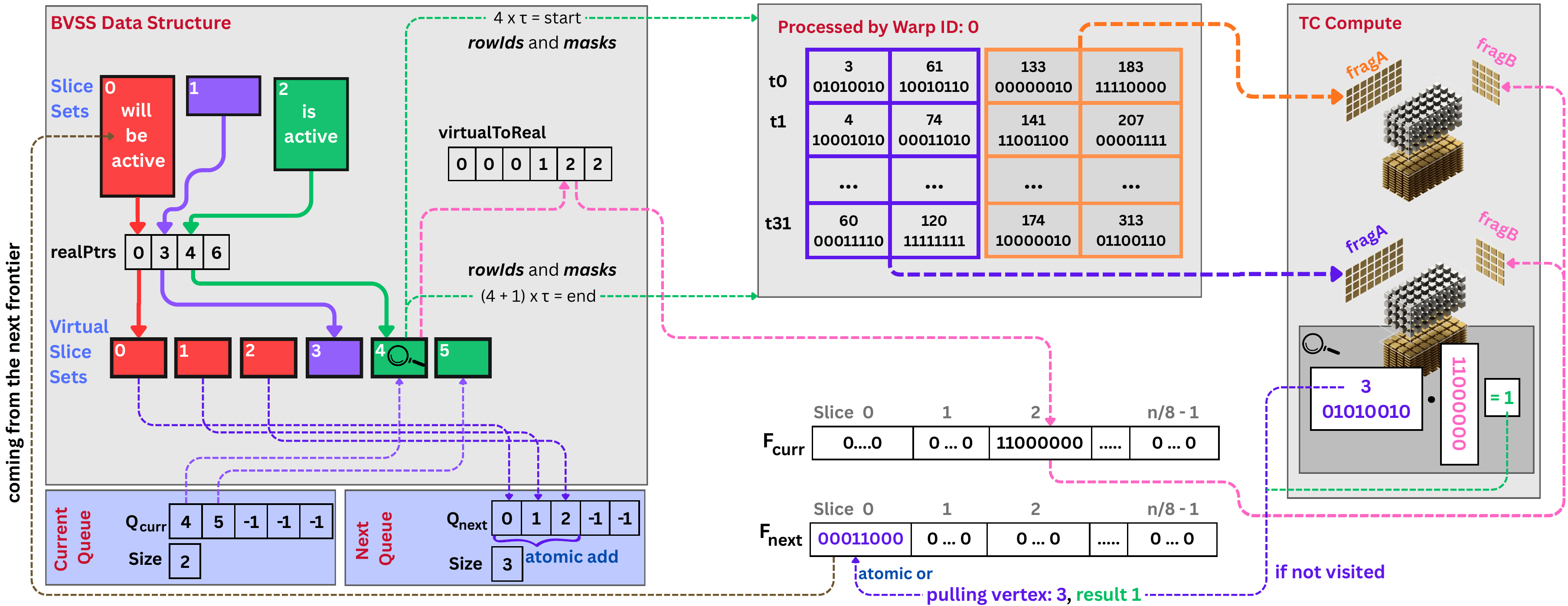}
    \caption{\small
      BVSS data structure and the flow of data reads, pull operations, and frontier updates:
      Slice set~2 (in green) is active, with \textbf{F}$_{\mathtt{curr}}$ bits {\tt 11000000}, i.e., the 1st and 2nd columns of slice set~2 are in the current frontier. Since \acro's queues operate over VSS indices, it retrieves the two VSSs corresponding to slice set~2, namely VSSs~4 and~5, directly from the current queue $\textbf{Q}_{\mathtt{curr}}$ (bottom left). For simplicity, the figure focuses on VSS~4 whose slices (${\mathtt{rowIDs}}$ and ${\mathtt{masks}}$) are read from memory and assigned Warp~0. The masks and the $\sigma$ frontier bits of slice set~2 (accessed via $\mathtt{virtualToReal}$ from VSS~4 and then from \textbf{F}$_{\mathtt{curr}}$) are fed to the TC as {\textit{fragA}} and {\textit{fragB}}, respectively. Pulls on the Boolean semiring are processed in two rounds, each processing half of the slices (colored purple and orange). For the slice with vertex/row~3 in the first half, the mask is {\tt 01010010}, indicating that vertex~3 is an outgoing neighbour of the 2nd, 4th, and 7th vertices/columns of slice set~2. Since the 2nd vertex is active in the frontier, the popcount is nonzero, and an update is required~(vertex~3 is unvisited). Since vertex~3 belongs to slice set~0 (covering vertices 0–7), its 4th bit is set in \textbf{F}$_{\mathrm{next}}$. Finally, \acro locates all VSSs corresponding to slice set~0 via $\mathrm{realPtrs}$ and inserts their IDs (0, 1, and~2) into \textbf{Q}$_{\mathrm{next}}$.}
     \label{fig:data_structure}
\end{figure*}

In this work, we decouple the bottom-up mechanism from pull-based exploration, as well as top-down from push-based, yielding four distinct modes of a parallel BFS. Although inefficient on CUDA cores, pull-based update mechanics can be practically coupled with a top-down work queue, thereby removing the assumption that these terms are interchangeable; in its simplest form, a thread responsible for updating a private state can go over all the frontier queue vertices and check their neighbourhoods to see if its state variable needs to be updated. Since pull-based exploration admits a matrix multiplication formulation suitable for TCs, we adopt it; however, we maintain a work queue containing all the frontier vertices (and some more), which is common for the top-down exploration in the state-of-the-art. 

The literature shows that switching the direction, and hence the update mechanism, during an iterative graph algorithm pays off well not only for a BFS but also for others~\cite{10.1145/3078597.3078616}. Adding the TCs to the list of available units yields another switching flexibility; an iteration can be processed not only by the CUDA cores but also by TCs. Furthermore, a hybrid processing is also possible, an avenue that has been explored for SpMM~\cite{hr_spmm, brp_spmm, hcspmm, rshspmm}. Hence, as shown in this work, existing BFS frameworks that successfully apply excellent switching mechanisms, yet use only CUDA cores, such as GSWITCH~\cite{gswitch}, may not fully utilize the recent GPUs. 

\section{\acro Data structure: BVSS}
\label{sec:ds}

The {\em Binarized Row Slice} (BRS) structure~\cite{berrybees} partitions $\mathbf{A}$ along its column dimension into sets of width $\sigma$, such that a nonzero $\mathbf{A}[i][j]$ falls in {\em slice set} $s$ if $s\sigma \leq j < (s+1)\sigma$. Row $i$ is included in slice set $s$ if it contains at least one nonzero in the column range $[s\sigma,\,(s+1)\sigma)$. Each such row is represented as a \emph{slice}, a row ID and a $\sigma$-bit mask, where $\mathtt{mask}[j - s\sigma]$ is set for every nonzero $\mathbf{A}[i][j]$. During computation, each slice set is assigned to a single warp, which assembles an intermediate matrix from the masks of that slice set 
and multiplies it against the corresponding partition $\mathbf{x}^{(k)}[s\sigma : (s+1)\sigma)$ of the current frontier vector. For a row $i$ processed, the warp sets $\mathtt{level}[i] = k+1$ and $x^{(k+1)}_i = 1$ if the mask-frontier multiplication produces a set bit, and $i$ has not been visited at any prior level. 
The SotA is inefficient due to four factors:
\begin{enumerate}[leftmargin=*, label=\textbf{\arabic*.}]
  \item Assigning one slice set per warp exposes an inter-warp load imbalance due to skewed degree distributions. 

 \item Slice sets are dispatched in a frontier-oblivious manner. The partition $\mathbf{x}^{(k)}[s\sigma : (s+1)\sigma]$ may be all zero, yet the assigned warp becomes active and must work. Even with early exits from warps due to all-zero frontiers, a warp may exit frequently while others do not.

  \item When the information ratio in a $\mathtt{mask}$, i.e., $\mathtt{popc}(\mathtt{mask})\,/\,\sigma$, is low, the TCs perform redundant operations. 

  \item The updates to $\mathtt{level}[i]$ and $x^{(k+1)}_i$ can cause high L1 cache miss rates and excessive L2 cache line fetches, respectively, due to scattered row IDs within a set.
\end{enumerate}

\noindent We address the first two problems in Sec.~\ref{subsec:bvss}, where we propose our novel data structure. We address the third one in Sec.~\ref{subsec:jaccard} and the last one in Secs.~\ref{subsec:rcm} and~\ref{subsec:lazy}.

\subsection{Binarized Virtual Slice Sets (BVSS)}
\label{subsec:bvss}

Let $\theta = 32/\sigma$ denote the number of slices assigned to each thread within a warp, and let $\tau = 32\theta$ denote the total number of slices constituting one unit of warp work. Motivated by this, we partition each slice set into a variable number of \emph{virtual slice sets}~(VSS), each carrying at most $\tau$ slices. The upper-left portion of Figure~\ref{fig:data_structure} illustrates this partitioning: the three slice sets are split into three, one, and two VSSs, respectively, implying that the first slice set contains $[2\tau + 1,\, 3\tau)$ slices, the third contains $[\tau + 1,\, 2\tau)$ slices, and the second contains fewer than $\tau$ slices. Let $N_s$ and $N_v$ denote the total number of slice sets and VSSs, respectively. The mapping from slice sets to their virtual children is maintained through an array $\mathtt{realPtrs}$ of size $N_s + 1$, where $\mathtt{realPtrs}[s+1] - \mathtt{realPtrs}[s]$ gives the number of VSSs emanating from slice set $s$. The inverse mapping is provided by $\mathtt{virtualToReal}$ of size $N_v$, where $\mathtt{virtualToReal}[v]$ identifies the parent slice set of VSS $v$. Throughout this work, we fix $\sigma = 8$, yielding $\theta = 4$ and $\tau = 128$. Consequently, each thread handles the multiplication of a 32-bit mask in collaboration with the rest of its warp, and is responsible for the update mechanism of 4 rows. VSSs containing fewer than $\tau$ slices---of which there are at most $N_s$ such sets, at most one per slice set---are zero-padded with all-zero masks and arbitrary row IDs to simplify indexing: both row IDs and masks are accessed at position $v\tau + \mathtt{laneID} \cdot \theta$, where $\mathtt{laneID}$ is the thread index within the warp processing virtual slice set $v$. Consequently, each $\mathtt{rowIds}$ element packs four 4-byte rows, and each $\mathtt{masks}$ element holds one 32-bit mask. No VSSs are constructed for slice sets that contain no slices. We name this data structure \textbf{Binarized VSSs (BVSS)}. 

Contrary to SotA~\cite{berrybees}, we maintain a work queue that contains only the VSSs that must be processed at the current BFS level, i.e., those whose parent slice set covers at least one column that belongs to the current frontier. In Fig.~\ref{fig:data_structure}, the third slice set is active, and all of its child VSSs $\{4,\,5\}$ reside in the queue simultaneously. Taking VSS $4$ as an example, the warp assigned to it resolves $\mathtt{virtualToReal}[4] = 2$ and accesses the corresponding 8-bit frontier partition \textbf{F}$^{\sigma}_{\mathtt{curr}}[2]$\footnote{
Throughout the paper, array subscripts indicate access granularity: for \(X\in\{\mathbf{F}_{\mathrm{curr}},\mathbf{F}_{\mathrm{next}},\mathbf{V}_{\mathrm{curr}},\mathbf{V}_{\mathrm{next}}\}\), \(X^1[u]\), \(X^\sigma[s]\), \(X^{32}[w]\), and \(X^\kappa[\cdot]\) denote vertex, slice-set, 32-bit-word, and \(\kappa\)-bit accesses, respectively.}. The multiplication of the $\tau = 128$ slices assigned to this warp is performed in two rounds of TC multiplications; the exact layout of the matrix constructed for each round is detailed in Sec.~\ref{subsec:layout}. When a multiplication yields a set bit, and the pulling vertex has never been visited before, as visualized in the bottom-right portion of Fig.~\ref{fig:data_structure}, \textbf{F}$^{\sigma}_{\mathtt{next}}[0]$, containing the bit for the pulling vertex $3$, is atomically updated. When a row within slice set $s$ is discovered for the first time, all VSSs in the range $[\mathtt{realPtrs}[s],\, \mathtt{realPtrs}[s+1])$ are enqueued into the next-level work queue. The algorithm terminates when the next-level queue is empty, i.e., when \textbf{F}$_{\mathtt{next}}$ is all-zero.

BVSS immediately resolves the first two problems stated in Sec.~\ref{sec:ds}. {\bf Inter-warp load balance} is resolved to near-perfect: any two warps selected from the grid can differ in the amount of work they perform by at most one virtual slice set, each containing a fixed and controlled number of slices. Such residual imbalance arises only when $|\mathbf{Q}_{\mathtt{curr}}| \bmod P \neq 0$, where $P$ is the number of warps in the grid, and is bounded by at most two TC multiplications across any two warps. The second deficiency, {\bf frontier-oblivious warp scheduling}, is also resolved. Warps are dispatched only for VSSs whose parent slice set covers at least one active frontier column. Consequently, every item submitted to the work queue corresponds to a genuinely active region of the graph through which BFS is transmitting information at the current level. 

\subsection{Top-Down Exploration with Pull-Based Mechanics}
\label{subsec:topdown_pull}

\begin{figure}[!t]
     \centering
     \includegraphics[width=0.5\textwidth]{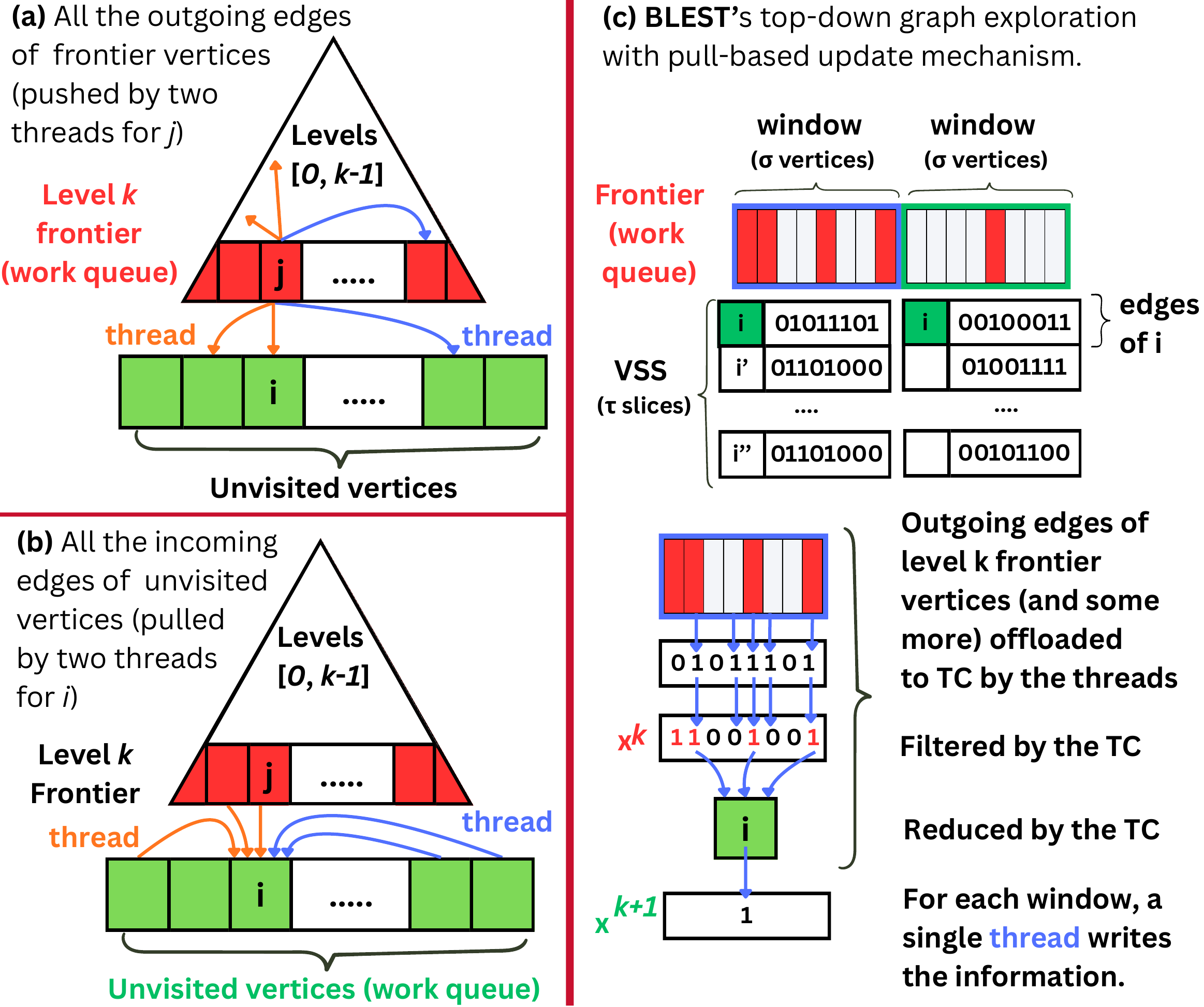}
    \caption{\small The traditional a) top-down~(push-based) and b) bottom-up~(pull-based)  BFS-level processing. c) \acro performs top-down exploration with a pull-based update mechanism.}
     \label{fig:pull_push}
\end{figure}

The work-queue dynamics of the pipeline depicted in Fig.~\ref{fig:data_structure} align with a top-down exploration; the work is determined by the vertices in the current frontier~(as in Fig.~\ref{fig:pull_push}a). To be exact, \acro processes the (outgoing) edges of the vertices that share a slice with at least one frontier vertex~(as in Fig.~\ref{fig:pull_push}c). However, it employs a pull-based approach; each thread is responsible for updating 4 (row) vertices. For each, the (incoming) neighbours determine whether the vertex resides in the frontier. In addition, since the (incoming) edges of a vertex $i$ may be scattered to multiple slice sets, multiple threads can update the state private to $i$.

\section{Reordering the Graph}
\label{sec:ordering}

To reduce the total number of slices, i.e., to compress BVSS, and to resolve locality issues during updates, \acro symmetrically reorders the graph $G$. We identify that graphs obeying power-law properties are highly compressible and introduce a scheme that increases the effective compression ratio by exploiting communities within slice sets. For graphs that are not scale-free-like, which are relatively less compressible, Reverse Cuthill--McKee (RCM)~\cite{rcm} is used to reduce the bandwidth of $\mathbf{A}$, thereby better confining the updates. 

\subsection{Increasing Compression Ratio}
\label{subsec:jaccard}

In BVSS, a slice exists within a set when the row it represents has at least one incoming neighbour among the $\sigma$ columns that the slice set covers. Consequently, the effective compression ratio of a slice with row ID $i$ residing in slice set $s$ is $\left(\sum_{j=0}^{\sigma-1} \mathbf{A}[i][s\sigma + j]\right) / \sigma$. To increase this ratio, one must construct slice sets such that the $\sigma$ columns extensively overlap among their outgoing row neighbours. 

Jaccard similarity~\cite{jaccard} is a popular approach to assess the overlaps between neighbourhoods.  
For two columns $j$ and $j'$, the similarity is defined as
\begin{equation}
    \mathcal{J}(j,\, j') = \frac{|\,\mathtt{nbrs}_{\mathbf{A}}(j) \cap \mathtt{nbrs}_{\mathbf{A}}(j')\,|}{|\,\mathtt{nbrs}_{\mathbf{A}}(j) \cup \mathtt{nbrs}_{\mathbf{A}}(j')\,|}.
    \label{eq:jaccard}
\end{equation}
Although computing all pairwise Jaccard values may produce desirable slice sets for BVSS, its complexity is $\mathcal{O}(n^2 \delta)$, where $\delta$ denotes the maximum column degree in $\mathbf{A}$, making it infeasible for real-world graphs. A basic observation for \acro is that only the $\sigma$ columns within each slice set need to form a community; inter-slice-set column affinity has no impact on compression. Motivated by this, we restrict the computation within {\em windows} of vertices and propose a novel reordering, \textsc{JaccardWithWindows}, given in Algorithm~\ref{alg:jaccard_with_windows}. 

The algorithm partitions $\mathbf{A}$ column-wise into non-overlapping $n/W$ windows of width $W > \sigma$ where $W \bmod \sigma = 0$. Each window contains $W/\sigma$ slice sets of size $\sigma$; columns are greedily assigned to slice sets one at a time. For each slice set, a highest-degree unassigned column within the window is selected as a singleton seed $j^*$, whose outgoing neighbour set initializes $\mathcal{R}$, with cost $\mathcal{O}(W)$~(line 6). The $\mathtt{inter}[j]$ values used to form the slice sets are initialized with cost $\mathcal{O}(\delta W)$~(line 9). 

The remaining $\sigma - 1$ slots are filled by repeatedly selecting the column $j \in \mathcal{Q}$---the set of unassigned columns within the window that share at least one row neighbour with $\mathcal{R}$---that maximizes $\mathtt{inter}[j] / (|\mathcal{R}| + \deg_{\mathbf{A}}(j) - \mathtt{inter}[j])$, where $\mathtt{inter}[j]$ is the number of common vertices within  $\mathtt{nbrs}_{\mathbf{A}}(j)$ and $\mathcal{R}$. When $\mathcal{Q}$ is empty, a highest-degree unassigned column is again chosen as a fallback~(line 13). Upon each selection, $\mathcal{R}$ is extended with the newly covered rows, and $\mathtt{inter}[\cdot]$ is incrementally updated via $\mathbf{A}^\top$ for all column candidates in $\mathcal{Q}$.  
Each slice set construction performs $\sigma$ scans of $\mathcal{Q}$ of size at most $W$~(lines 12--15). For the $\mathtt{inter}[\cdot]$ updates, each selected column introduces $\delta$ new rows into $\mathcal{R}$, and each such row contributes at most $\min(\delta, W)$ column candidate updates. Hence, the update cost is 
\begin{align}
\mathcal{O}\left(\frac{n}{W} \times \frac{W}{\sigma} \times \left(\delta W +  \sigma(W + \delta\min(\delta, W)\right)\right) \\
= \mathcal{O}\left(n \times \left(\frac{\delta W}{\sigma} + W + \delta\min(\delta, W)\right)\right)
\end{align}
in total. With $W \ll n$, the overall complexity is substantially below the $\mathcal{O}(n^2\delta)$ of the naive implementation. We evaluate the window size $W$ in Section~\ref{sec:experiments} and how it affects both the execution time and the compression rate of the algorithm.

\begin{algorithm}[!t]
\small
\caption{\textsc{JaccardWithWindows}}
\label{alg:jaccard_with_windows}
\begin{algorithmic}[1]
\Require{$\mathbf{A} \leftarrow G^\top$, $\mathbf{A}^\top \leftarrow G$, slice size $\sigma$, window size $W$}
\Ensure{Inverse permutation $\pi^{-1}$}
\For{$w = 0, \ldots, \lceil n/W \rceil - 1$ \textbf{in parallel}}
    \State $w_s \leftarrow wW$; \quad $w_e \leftarrow \min(w_s + W,\, n)$
    \State $\mathcal{C} \leftarrow \emptyset$
    \For{$s = 0, \ldots, (w_e - w_s)/\sigma - 1$}
        \State $s_s \leftarrow w_s + s\sigma$; \quad $s_e \leftarrow \min(s_s + \sigma,\, w_e)$
        \State $j^* \leftarrow \arg\max_{j \in [w_s,\, w_e) \setminus \mathcal{C}}\, \deg_{\mathbf{A}}(j)$
        \State $\mathcal{C} \leftarrow \mathcal{C} \cup \{j^*\}$; \quad $\pi^{-1}[j^*] \leftarrow s_s$
        \State $\mathcal{R} \leftarrow \mathtt{nbrs}_{\mathbf{A}}(j^*)$
        \State $\forall\, j \in [w_s, w_e) \setminus \mathcal{C}:\; \mathtt{inter}[j] \leftarrow |\mathcal{R} \cap \mathtt{nbrs}_{\mathbf{A}}(j)|$
        \State $\mathcal{Q} \leftarrow \{j \in [w_s, w_e) \setminus \mathcal{C} : \mathtt{inter}[j] > 0\}$
        \For{$\ell = s_s + 1, \ldots, s_e - 1$}
            \If{$\mathcal{Q} = \emptyset$}
                \State $j^* \leftarrow \arg\max_{j \in [w_s,\, w_e) \setminus \mathcal{C}}\, \deg_{\mathbf{A}}(j)$
            \Else
                \State $j^* \leftarrow \arg\max_{j \in \mathcal{Q}}\, \dfrac{\mathtt{inter}[j]}{|\mathcal{R}| + \deg_{\mathbf{A}}(j) - \mathtt{inter}[j]}$
            \EndIf
            \State $\mathcal{C} \leftarrow \mathcal{C} \cup \{j^*\}$; \quad $\mathcal{Q} \leftarrow \mathcal{Q} \setminus \{j^*\}$; \quad $\pi^{-1}[j^*] \leftarrow \ell$
            \For{$i \in \mathtt{nbrs}_{\mathbf{A}}(j^*) \setminus \mathcal{R}$}
                \State $\mathcal{R} \leftarrow \mathcal{R} \cup \{i\}$
                \For{$j \in \mathtt{nbrs}_{\mathbf{A}^\top}(i) \cap ([w_s, w_e) \setminus \mathcal{C})$}
                    \If{$\mathtt{inter}[j] = 0$}
                        \State $\mathcal{Q} \leftarrow \mathcal{Q} \cup \{j\}$
                    \EndIf
                    \State $\mathtt{inter}[j] \mathrel{+}= 1$
                \EndFor
            \EndFor
        \EndFor
    \EndFor
\EndFor
\State \Return $\pi^{-1}$
\end{algorithmic}
\end{algorithm}

\subsection{Boosting Cache Locality}
\label{subsec:rcm}

In a GPU, each SM has a private L1 cache, whereas the L2 cache is shared among all SMs. All atomic stores are therefore coherent in L2, enabling concurrent memory updates to be visible across all SMs. When a TC multiplication produces a nonzero result for row/vertex $i$ that has never been visited before, two updates indexed by row ID $i$ occur: (i) $\mathtt{level}[i] = k$, and (ii) \textbf{F}$^1_{\mathtt{next}}[i] \overset{\mathtt{atomic}}{\leftarrow} 1$. The efficiency of these updates depends entirely on the distribution of row IDs within a VSS: to have high L1 and L2 cache hit rates, respectively, for $\mathtt{level}[i]$ updates and \textbf{F}$^1_{\mathtt{next}}[i]$ atomic stores, the row IDs constituting a VSS must be tightly clustered. For graphs that are not scale-free-like such as road networks, 
reordering $G^\top$ so that rows within a VSS are highly local is both feasible and effective. To quantify this locality, we introduce a metric called \emph{update divergence}. For each of the $\theta$ columns of the VSS matrix shown in Fig.~\ref{fig:data_structure}, we define the \emph{column divergence} as the standard deviation of the row IDs with nonzero masks. 
A higher standard deviation indicates that the row IDs are more scattered and less clustered, whereas a lower value reflects tighter locality. The \emph{set divergence} of VSS $s$, denoted $\mathcal{U}_{\mathtt{div}}(s)$, is then the average column divergence over its non-empty columns. Finally, the \emph{update divergence} is the mean $\mathcal{U}_{\mathtt{div}} = \underset{s}{\mathtt{avg}}\;\mathcal{U}_{\mathtt{div}}(s)$ computed over all $N_v$ VSSs, where a lower $\mathcal{U}_{\mathtt{div}}$ indicates better locality.

\begin{figure*}[htbp]
     \centering
     \includegraphics[width=\textwidth]{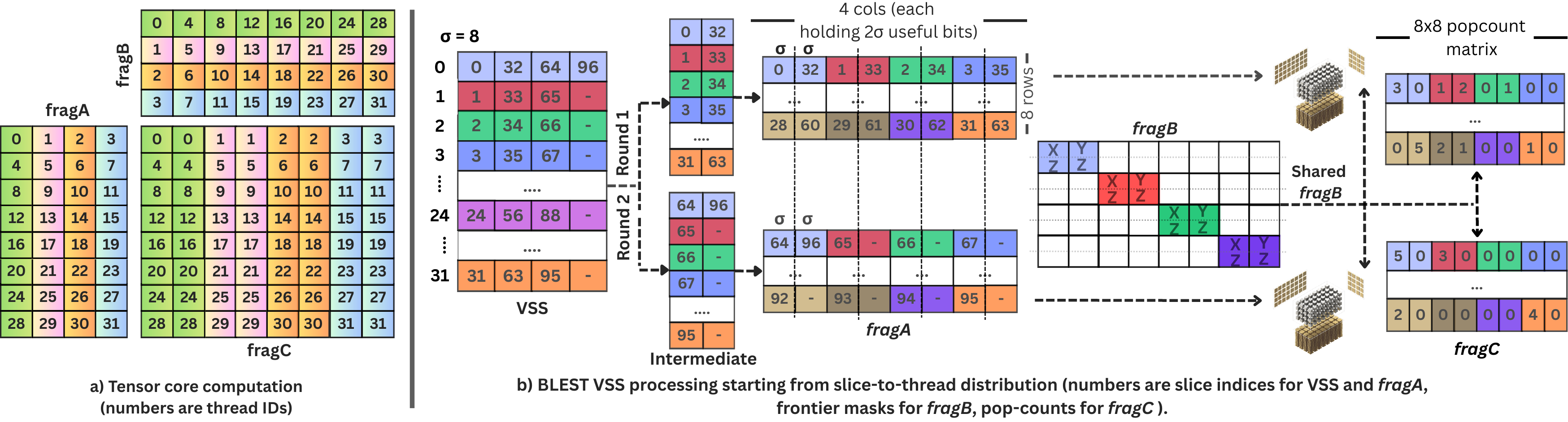}
     \caption{\small (a) The data layout for \texttt{m8n8k128} on Tensor Cores. For \texttt{fragA}, \texttt{fragB}, and \texttt{fragC}, each box corresponds to a 32-bit word, and the number inside is the ID of the thread holding that word in its registers. (b) \acro's VSS processing pipeline starting from the $32 \times 4$ VSS matrix (numbers denote slice indices for the VSS and \texttt{fragA}, frontier masks for \texttt{fragB}, and popcounts for \texttt{fragC}). The pull operations for a VSS are handled in two rounds, each issuing one \texttt{m8n8k128} instruction. In each round, two columns of the $32 \times 4$ VSS matrix are reshaped into a $32 \times 2$ intermediate matrix, then conceptually into the $8 \times 8$ \texttt{fragA} matrix (dashed lines), and finally packed into the $8 \times 4$ \texttt{fragA} matrix, where each 32-bit entry holds two $\sigma$-bit masks in its first $2\sigma$ bits. A single $4 \times 8$ \texttt{fragB} is shared across both rounds: letting $\alpha$ be the $\sigma$-bit frontier word and $\vartheta$ the all-zero $\sigma$-bit word, the 16-bit building blocks are $X = \alpha\vartheta$, $Y = \vartheta\alpha$, and $Z = \vartheta\vartheta$. All \texttt{fragB} words are zero except those held by threads $0, 4, 9, 13, 18, 22, 27$, and $31$, which carry $XZ$ or $YZ$ to select the first or second packed slice mask, respectively. This layout is \underline{optimal}: no \texttt{fragC} popcount is wasted, and every result is delivered directly to the thread that owns the corresponding slices, requiring no intra-warp communication.}
    \label{fig:multiplication_layout}
\end{figure*}

Reverse Cuthill-McKee (RCM)~\cite{rcm} 
assigns vertex IDs based on their visit order in a BFS-like traversal; vertices with the same parent are ordered by ascending degree, delaying frontier explosion and keeping connected vertices close. 
Table~\ref{tab:locality_values} reports $\mathcal{U}_{\mathtt{div}}$ for four graphs---two road networks~({\tt{GAP-road}} and {\tt{europe\_osm}}), a planar, triangulation network ({\tt{delaunay\_n24}}), and one random network~({\tt{rgg\_24}})---before and after RCM. For all, the dramatic reduction in $\mathcal{U}_{\mathtt{div}}$ confirms that RCM substantially tightens the clustering of row IDs within VSSs.

\begin{table}[h]
\centering
\scalebox{0.77}{
\begin{tabular}{lrr|lrr}\\  \\
  \textbf{Graph} &
  \textbf{\begin{tabular}[r]{@{}r@{}}$\mathcal{U}_{\mathtt{div}}$\\(unord.)\end{tabular}} &
  \textbf{\begin{tabular}[r]{@{}r@{}}$\mathcal{U}_{\mathtt{div}}$\\(ord.)\end{tabular}} & \textbf{Graph} &
  \textbf{\begin{tabular}[r]{@{}r@{}}$\mathcal{U}_{\mathtt{div}}$\\(unord.)\end{tabular}} &
  \textbf{\begin{tabular}[r]{@{}r@{}}$\mathcal{U}_{\mathtt{div}}$\\(ord.)\end{tabular}} \\
  \hline
  \tt{GAP-road}      & 158{,}696 & 869 & \tt{europe\_osm}   & 869{,}302 & 982  \\
  \tt{delaunay\_n24} & 894{,}542 & 2{,}839 &  \tt{rgg\_24}       &   3{,}948 & 1{,}512 \\
  \hline
\end{tabular}
}
\caption{\small Avg. update divergence $\mathcal{U}_{\mathtt{div}}$ before and after RCM.}
\label{tab:locality_values}
\end{table}

Unfortunately, RCM is not effective for scale-free-like graphs, which do not generally possess a recoverable low bandwidth. During preprocessing, we therefore first identify whether $G$ is scale-free-like\footnote{We classify a graph as scale-free-like if either its in- or out-degree distribution is heavy-tailed: the top \(1\%\) and \(10\%\) of vertices account for at least \(5\%\) and \(40\%\) of total degree, respectively, or a log-log degree-histogram fit for \(k\ge5\) has slope \(-\gamma\), \(R^2\ge0.70\), and \(\gamma\in[1,5]\).}. If so, we apply \textsc{JaccardWithWindows} (Section~\ref{subsec:jaccard}), which aims to maximize mask density within each slice set. Otherwise, we apply RCM to reorder $G^\top$, which reduces $\mathcal{U}_{\mathtt{div}}$ and thereby improves cache efficiency during warp-level updates. Although $\mathcal{U}_{\mathtt{div}}$ measures per-virtual-slice-set locality, RCM also improves intra-virtual-slice-set locality: row IDs assigned to adjacent virtual slice sets are numerically proximate, so warps processing neighbouring VSSs also operate on clustered row IDs. Specifically, the $\mathtt{level}[i]$ update benefits when those warps are co-located on the same SM, since the relevant cache lines are then likely to reside in the shared L1 cache; the \textbf{F}$^1_{\mathtt{next}}[i]$ atomic update benefits unconditionally, as the L2 cache is shared across all SMs and clustered row IDs translate directly to higher L2 cache hit rates regardless of warp placement.

\section{Compute Mechanics of \acro}
\label{sec:compute}

As stated before, mapping the edge operations within a BFS onto TCs is non-trivial. Specifically, the matrix--vector primitive that underpins single-source BFS must be recast as a specialized form of matrix--matrix multiplication that these dense units are built for. Even a layout that efficiently achieves such a recasting is not sufficient to extract peak hardware performance from a BFS kernel, which is widely regarded as a memory-bound algorithm. To address both concerns, we first introduce our novel TC multiplication layout in Sec.~\ref{subsec:layout}, which {\em optimally} maps bit sparse-matrix--sparse-vector multiplication onto the dense units, and then in Sec.~\ref{subsec:lazy}, we propose a lazy vertex-update scheme that substantially alleviates the inherent memory inefficiencies.

\subsection{An Optimal Layout for TC Multiplication}
\label{subsec:layout}

Fig.~\ref{fig:multiplication_layout} depicts the TC multiplication layout used in \acro. In each of the two rounds per VSS, the warp processes 2/4 columns of the $32 \times 4$ VSS matrix. These two columns form a $32 \times 2$ intermediate matrix, conceptually reshaped into an $8 \times 8$ \texttt{fragA} matrix delimited by dashed lines in part~(b). Since the \texttt{m8n8k128} TC instruction expects \texttt{fragA} to have shape $8 \times 4$, two slices are packed into each \texttt{fragA} entry: for every row of the $8 \times 8$ \texttt{fragA} matrix, the two $\sigma$-bit connectivity patterns belonging to the two selected columns are concatenated into the first $2\sigma$ bits of a 32-bit element, with the remaining 16 bits set to zero, reducing the $8 \times 8$ \texttt{fragA} matrix to the $8 \times 4$ \texttt{fragA} matrix. This packing is optimal: the $8 \times 4$ \texttt{fragA} matrix cumulatively holds data for exactly 64 slices, so multiplying against a single $\sigma$-bit frontier word fills all 64 entries of the $8 \times 8$ \texttt{fragC} matrix with useful popcounts, wasting no output. \acro uses a single $4 \times 8$ \texttt{fragB} shared across both rounds, as shown in part~(b). Let $\alpha$ denote the $\sigma$-bit frontier word and $\vartheta$ the all-zero $\sigma$-bit word. We define three 16-bit building blocks: $X = \alpha\vartheta$, $Y = \vartheta\alpha$, and $Z = \vartheta\vartheta$. Each row $i \in \{0, 1, 2, 3\}$ of \texttt{fragB} contains exactly two nonzero 32-bit words, located at columns $j = 2i$ and $j = 2i + 1$; all remaining entries are zero. For the eight nonzero entries, if $j$ is even, the corresponding 32-bit word is $XZ$, and if $j$ is odd, it is $YZ$, as visualized in the \texttt{fragB} panel of part~(b). Consequently, for even $j$, \texttt{fragB} selects the first packed slice via $X$, and for odd $j$ it selects the second via $Y$, while the $Z$ components mask out the unused halves.

To avoid intra-warp communication for accumulation, we exploit the distribution of \texttt{fragB} words to warp threads, as in part~(a) of Fig.~\ref{fig:multiplication_layout}. The 32-bit words of \texttt{fragB} are all zeros except those held by threads satisfying $t \bmod 9 = 0$, which hold $XZ$, and threads satisfying $t \bmod 9 = 4$, which hold $YZ$. This distribution is unique: thread $t$ owns the \texttt{fragC} entries at $(i, j)$ and $(i, j+1)$, where $i = \lfloor t/4 \rfloor$ and $j = 2 \times (t \bmod 4)$, and $t$'s packed slices reside in $\texttt{fragA}[i][j/2]$. That \texttt{fragA} entry is multiplied by $\texttt{fragB}[j/2][j] = XZ$ and $\texttt{fragB}[j/2][j+1] = YZ$, which select the first and second slice masks of $t$ respectively, writing their outputs directly to $\texttt{fragC}[i][j]$ and $\texttt{fragC}[i][j+1]$---the very entries owned by $t$. Consequently, every thread retrieves its own slice--frontier dot-product results from its own \texttt{fragC} registers with no intra-warp communication while reducing the number of required MMA calls $8\times$ compared to~\cite{berrybees}.
The basic \acro algorithm based on our novel layout is summarized in Alg.~\ref{alg:bvss_kernel}.

\algnewcommand{\IIf}[1]{\State\algorithmicif\ #1\ \algorithmicthen}
\algnewcommand{\EndIIf}{\unskip\ }
\algrenewcommand\algorithmicindent{0.9em}%
\begin{algorithm}[t]
\small
\caption{\acro{}}
\label{alg:bvss_kernel}
\begin{algorithmic}[1]
\Require (1) BVSS data structure, (2) $src$: source vertex
\Ensure (1) \(\textbf{L}\): level array
\State \(\textbf{L}[v] \gets \infty,\ \forall v \in \mathcal{V}/\{src\}\); \(\textbf{L}[src] \gets 0\); \Comment{init}
\State \({\textnormal{\textbf{F}}}^1_{\text{curr}}[v] \gets 0,\ \forall v \in \mathcal{V}/\{src\}\); \({\textnormal{\textbf{F}}}^1_{\text{curr}}[src] \gets 1\); \Comment{init}
\State \({\textnormal{\textbf{F}}}^n_{\text{next}}[v] \gets 0\); \Comment{init}
\State $\textbf{Q}_{\text{curr}} \leftarrow [\textit{\textbf{realPtrs}}[\lfloor src/\sigma\rfloor],\ \textit{\textbf{realPtrs}}[\lfloor src/\sigma\rfloor+1])$; \Comment{init}
\State $\textbf{Q}_{\text{next}} \leftarrow \emptyset$
\State \(\ell \gets 0\); \Comment{current BFS level}
\State \(\textit{cont} \gets \texttt{\footnotesize true}\); \Comment{non-empty state of the frontier}
\While{\(\textit{cont}\)}
  \State \(\ell \gets \ell + 1\);
  \For{\(w = \textit{warpID};\ w < |\textbf{Q}_{\text{curr}}|;\ w \gets w + \#\text{warps}\)}
    \State \(vss_{in} \gets \textbf{Q}_{\text{curr}}[w]\); \Comment{virtual slice set ID}
    \State \(ss_{in} \gets \textit{\textbf{virtualToReal}}[vss_{in}]\); \Comment{slice set ID}
    \State \(\textit{tile} \gets (vss_{in} \ll 5) + \textit{laneID}\);
    \State \((u_0,u_1,u_2,u_3) \gets \textit{\textbf{rowIds}}[\textit{tile}]\); \Comment{vectorized 128-bit read}
    \State \(\textit{mask} \gets \textit{masks}[\textit{tile}]\); \Comment{$4 \times \sigma = 32$ bit read}
    \State \(\alpha \gets \textbf{F}^{\sigma}_{\text{curr}}[ss_{in}]\); \Comment{$\sigma$-bit frontier word}
    \State \(\textit{fragB} \gets 0\);
    \IIf{\(\textit{laneID} \bmod 9 = 0\)} \(\textit{fragB} \gets \alpha\); \EndIIf
    \IIf{\(\textit{laneID} \bmod 9 = 4\)} \(\textit{fragB} \gets \alpha\); \(\textit{fragB} \gets \textit{fragB} \ll 8\); \EndIIf
    \State \(\textit{fragA} \gets \text{low16}(\textit{mask})\); \Comment{the first two masks}
    \State \(\textit{fragC}[0,1] \gets \mathtt{TC}(\textit{fragA}, \textit{fragB})\); \Comment{1st \texttt{m8n8k128}}
    \State \(\textit{fragA} \gets \text{high16}(\textit{mask})\); \Comment{the last two masks}
    \State \(\textit{fragC}[2,3] \gets \mathtt{TC}(\textit{fragA}, \textit{fragB})\); \Comment{2nd \texttt{m8n8k128}}
    \For{\(c \in \{0,1,2,3\}\)} \Comment{column in VSS matrix}
      \If{\(\textit{fragC}[c] \neq 0\)} \Comment{if dot-product is nonzero}
        \State \(u \gets (u_0,u_1,u_2,u_3)[c]\); \Comment{row ID to update}
        \State \(\ell_{\text{prev}} \gets \textbf{L}[u]\);
        \If{\(\ell < \ell_{\text{prev}}\)} \Comment{if $u$ is not visited}
          \State \(\textbf{L}[u] \gets \ell\); \Comment{set $u$'s level}
          \State \(ss_{out} \gets \lfloor u/\sigma \rfloor\); \Comment{$u$'s slice set index}
          \State \(old \gets \textbf{F}^{\sigma}_{\text{next}}[ss_{out} ] \stackrel{\text{atomic}}{\lor} (1 \ll (u \bmod \sigma)) \); 
          \If{\(old = 0\)} \Comment{$u$'s set is seen for the first time}
            \State \([s,e) \gets [\textit{\textbf{realPtrs}}[ss_{out}],\ \textit{\textbf{realPtrs}}[ss_{out}+1])\);
            \State \(\textbf{Q}_{\text{next}} \stackrel{\text{atomic}}{\gets} \textbf{Q}_{\text{next}} \cup [s,e)\);
          \EndIf
        \EndIf
      \EndIf
    \EndFor
  \EndFor
  \State \textsc{GridSync}(); \Comment{level synchronization}
  \State \(\textit{cont} \gets (|\textbf{Q}_{\text{next}}| > 0)\); \Comment{check the frontier state}
  \State \(\text{swap}(\textbf{F}_{\text{curr}}, \textbf{F}_{\text{next}})\); \(\text{swap}(\textbf{Q}_{\text{curr}}, \textbf{Q}_{\text{next}})\); \Comment{swap arrays}
  \State \textsc{GridSync}();
  \State \(|\textbf{Q}_{\text{next}}| \gets 0\); \(\textbf{F}^n_{\text{next}} \gets 0\); \Comment{clear next frontier data}
  \State \textsc{GridSync}();
\EndWhile
\end{algorithmic}
\end{algorithm}

\subsection{Lazy Vertex Updates}
\label{subsec:lazy}

When the graph bandwidth and update divergence are high and cannot be reduced by reordering, lines 27, 29, and 31 of Alg.~\ref{alg:bvss_kernel} suffer from poor L1/L2 cache hit rates, respectively, as the row IDs updated per warp are highly scattered. Beyond locality, two additional factors degrade performance. First, lines 31 and 34 each issue an atomic operation in the hot path, stalling the warp until the operation completes in L2 cache and its return value is received. Second, lines 27/28 introduce a race condition: multiple threads may concurrently read the same $u$, pass the unvisited check, and proceed to line 31, inflating the number of atomic operations. The atomic at line 31 correctly prevents duplicate queue insertions, but at the cost of amplifying atomic contention. RCM partially mitigates the race at lines 27/28 for networks that are not scale-free-like by co-scheduling warps on the same SM over clustered, largely overlapping row ID ranges, increasing the likelihood that a concurrent level update is already visible in the shared L1 cache before a second thread checks. Furthermore, on the same networks, the per-level frontier is small enough that the absolute number of atomics remains manageable. On scale-free-like networks, however, their cumulative effect is detrimental. To address  these bottlenecks simultaneously, we introduce the \emph{lazy vertex update} scheme.

The GPU instruction set architecture (SASS) provides two instructions that ensure atomicity: \textbf{ATOMG} and \textbf{REDG}. Although both guarantee full atomicity, \textbf{REDG} is the asynchronous counterpart of \textbf{ATOMG}: the compiler emits \textbf{REDG} whenever it detects that the return value of an atomic operation is not used by the calling warp, allowing the warp to continue execution without waiting for the previous memory value to be returned---a capability that \textbf{ATOMG} does not offer. Motivated by this, the lazy vertex update scheme defers all necessary updates to the end of the current BFS level by replacing the synchronous atomic operations in Algorithm~\ref{alg:bvss_kernel} with their asynchronous \textbf{REDG} counterparts, and resolves them fully only once the entire BFS level completes. The modified algorithm is provided in Algorithm~\ref{alg:bvss_kernel_lazy}.

\algrenewcommand\algorithmicindent{1.1em}%
\begin{algorithm}[t]
\small
\caption{\acro{} with Lazy Vertex Updates}
\label{alg:bvss_kernel_lazy}
\begin{algorithmic}[1]
\Require (1) BVSS data structure, (2) $src$: source vertex
\Ensure (1) \(\textbf{L}\): level array
\State \(\textbf{L}[v] \gets \infty,\ \forall v \in \mathcal{V}/\{src\}\); \(\textbf{L}[src] \gets 0\); \Comment{init}
\State \({\textnormal{\textbf{F}}}^1_{\text{curr}}[v] \gets 0,\ \forall v \in \mathcal{V}/\{src\}\); \({\textnormal{\textbf{F}}}^1_{\text{curr}}[src] \gets 1\); \Comment{init}
\State \({\textnormal{\textbf{V}}}^1_{\text{curr}}[v] \gets 0,\ \forall v \in \mathcal{V}/\{src\}\); \({\textnormal{\textbf{V}}}^1_{\text{curr}}[src] \gets 1\); \Comment{init}
\State \({\textnormal{\textbf{V}}}^1_{\text{next}}[v] \gets 0,\ \forall v \in \mathcal{V}/\{src\}\); \({\textnormal{\textbf{V}}}^1_{\text{next}}[src] \gets 1\); \Comment{init}
\State $\textbf{Q}_{\text{curr}} \leftarrow [\textit{\textbf{realPtrs}}[\lfloor src/\sigma\rfloor],\ \textit{\textbf{realPtrs}}[\lfloor src/\sigma\rfloor+1])$;  \Comment{init}
\State \(\ell \gets 0;\) \Comment{current BFS level}
\State \(\textit{cont} \gets \texttt{true};\) \Comment{non-empty state of the frontier}
\While{\(\textit{cont}\)}
  \State \(\ell \gets \ell + 1;\)
  \State \textbf{Stage 1: Lazy marking\ \hrulefill}
  \For{\(w = \textit{warpID};\ w < \lvert \textbf{Q}_{\text{curr}} \rvert;\ w \gets w + \#\text{warps}\)}
  
    \State \hspace*{0.4ex}... \\ \hspace*{5.2ex}Lines 11--23 of Algorithm~\ref{alg:bvss_kernel} \\ \hspace*{5.2ex}...
    \setcounter{ALG@line}{25}
    
    \For{\(c \in \{0,1,2,3\}\)} \Comment{column in VSS matrix}
      \If{\(\textit{fragC}[c] \neq 0\)} \Comment{if dot-product is nonzero}
        \State \(u \gets (u_0,u_1,u_2,u_3)[c];\) \Comment{row ID to update}
        \State \({\textnormal{\textbf{V}}}^1_{\text{next}}[u] \;\stackrel{\text{atomic}}{\lor}\; 1;\) \Comment{lazy mark}
      \EndIf
    \EndFor
  \EndFor
  \State \(\lvert \textbf{Q}_{\text{next}} \rvert \gets 0;\) \Comment{prepare for Stage 2}
  \State \textsc{GridSync}();
  \vspace*{1ex}
  \State \textbf{Stage 2: Frontier finalization\ \hrulefill}
  \vspace*{0.7ex}
  \For{\(t = \textit{threadID};\ t < \lceil{n/32}\rceil;\ t \gets t + \#\text{threads}\)}
    \State \(\textit{next} \gets {\textnormal{\textbf{V}}}^{32}_{\text{next}}[t];\) \Comment{vertices visited until here}
    \State \(\textit{diff} \gets {\textnormal{\textbf{V}}}^{32}_{\text{curr}}[t] \,\mathbin{{\oplus}}\, \textit{next};\) \Comment{new frontiers in Stage~1}
    \State \(\textit{rssOffset} \gets 4t;\) \Comment{4 slice sets per word}
    \If{\(\textit{diff} \neq 0\)} \Comment{next frontier is non-empty}\vspace*{0.7ex}
      \State \({\textnormal{\textbf{V}}}^{32}_{\text{curr}}[t] \gets \textit{next};\)\vspace*{0.7ex}
      \State \({\textnormal{\textbf{F}}}^{32}_{\text{curr}}[t] \gets \textit{diff};\) \Comment{set bits for frontier vertices}
      \For{\(\textit{set} \in \{0,1,2,3\}\)} \Comment{each 8-bit word in $diff$}
        \State \(ss_{mask} \gets (\textit{diff} \gg (8 \cdot \textit{set})) \mathbin{\&}\ 0x\text{FF};\) \Comment{select 8-bits}
        \If{\(ss_{mask} \neq 0\)} \Comment{change is in this set}
          \State \(ss_{out} \gets \textit{rssOffset} + \textit{set};\)
          \While{\(ss_{mask} \neq 0\)} \Comment{go over all bits}
            \State \(b \gets \textbf{\_ffs}(ss_{mask}) - 1;\) \Comment{in \(\{0,\dots,7\}\)}
            \State \(u \gets ss_{out} \cdot \sigma + b;\) \Comment{new frontier vertex}
            \State \(\textbf{L}[u] \gets \ell;\)
            \State \(ss_{mask} \gets ss_{mask} \mathbin{\&} (ss_{mask} - 1);\) \Comment{unset $u$}
          \EndWhile
          \State \([s,e) \gets [\textit{\textbf{realPtrs}}[ss_{out}],\ \textit{\textbf{realPtrs}}[ss_{out}+1]);\)
          \State \(\textbf{Q}_{\text{next}} \;\stackrel{\text{warp-atomic}}{\gets}\; \textbf{Q}_{\text{next}} \;\cup\; [s,e);\)
        \EndIf
      \EndFor
    \EndIf
  \EndFor
  \State \textsc{GridSync}();
  \State \(\textit{cont} \gets (\lvert \textbf{Q}_{\text{next}} \rvert > 0);\)
  \State \(\text{swap}(\textbf{Q}_{\text{curr}}, \textbf{Q}_{\text{next}});\)
  \State \textsc{GridSync}();
\EndWhile
\end{algorithmic}
\end{algorithm}

Alg.~\ref{alg:bvss_kernel_lazy} introduces two additional arrays, $\mathbf{V}_{\text{curr}}$ and $\mathbf{V}_{\text{next}}$, holding the cumulative visit status of all vertices up to and including the current and next BFS levels, respectively. The set difference $\mathbf{V}_{\text{next}} \setminus \mathbf{V}_{\text{curr}}$ therefore identifies exactly the vertices joining the frontier at the next level. The per-level computation is split into two stages.

\textbf{Stage~1 (Lazy marking).} The TC multiplication proceeds identically to Alg.~\ref{alg:bvss_kernel}, but for every {\em pulling} vertex $u$ that produces a nonzero dot-product, $\mathbf{V}^1_{\text{next}}[u]$ is marked lazily via an asynchronous \textbf{REDG} atomic {\bf{or}}, with no return value consumed. Crucially, not every such vertex genuinely requires an update: vertices already visited at prior levels are filtered out in Stage~2. Deferring this decision eliminates all synchronous \textbf{ATOMG} instructions from the hot path.

\textbf{Stage~2 (Frontier finalization).} Once all warps complete Stage~1, a \textsc{GridSync} separates the two stages, after which all threads sweep $\mathbf{V}_{\text{next}}$ collectively in a fully coalesced manner. Each thread processes one 32-bit word per iteration, computing $\textit{diff} = \mathbf{V}^{32}_{\text{curr}}[t] \oplus \mathbf{V}^{32}_{\text{next}}[t]$ to isolate the bits corresponding to vertices that are genuinely new to the frontier. The $\sigma$-bit partition $ss_{mask}$ of $\textit{diff}$ for each of the 4 slice sets in the word is extracted and, if nonzero, iterated via the \textbf{\_ffs} instruction---which locates the lowest set bit in a single cycle---to identify each new frontier vertex $u$, assign $\mathbf{L}[u] \gets \ell$, and enqueue the corresponding VSS range $[\mathtt{realPtrs}[ss_{out}],\, \mathtt{realPtrs}[ss_{out}+1])$. Since threads are assigned to disjoint words, the $\mathbf{F}^{32}_{\text{curr}}[t]$ and $\mathbf{V}^{32}_{\text{curr}}[t]$ updates require no atomics, contrary to Alg.~\ref{alg:bvss_kernel}, where $\mathbf{F}_{\text{curr}}$ updates require atomics with expensive cache costs. The level updates $\mathbf{L}[u] \gets \ell$, on the other hand, exploit high spatial locality since $u$ values within a word are consecutive. Furthermore, the VSS enqueuing is performed warp-atomically, making the only full atomic part of the algorithm immensely cheap, reducing the required number of atomic operations by a factor of $32\times$ relative to only this part of Alg.~\ref{alg:bvss_kernel}. Although Stage~2 introduces a $\Theta(n)$ sweep per BFS level, we identify that on scale-free-like networks with only tens of BFS levels, it removes almost the entire atomic bottleneck and resolves cache problems, making the trade-off highly beneficial. 

\subsection{Switching Mechanism of \acro}
\label{subsec:switching}

During a BFS, a frontier crowded with many VSSs may become a bottleneck even with the optimizations described above. In such cases, bottom-up exploration of the next frontier with pull-based mechanics on CUDA cores may be preferable. 
Indeed, the graphs that may benefit from this are those for which lazy updates are useful.  
Conveniently, the 2nd stage, {\em frontier finalization} described above, traverses the entire $\mathbf{V}_{\text{curr}}$, and the number of unvisited vertices can be computed as the popcount of its bitwise complement. This value, $\#unvisited$, is used by \acro to decide whether to switch at the start of each level based on the condition
\begin{equation}
    \#unvisited < \eta \cdot |\textbf{Q}_{\text{curr}}|,
    \label{eq:switching}
\end{equation}
where $|\textbf{Q}_{\text{curr}}|$ is the current frontier queue size, and $\eta$ is the switching constant, set to $10$ based on our preliminary results. When the inequality~\eqref{eq:switching} is met for a level, \acro invokes Alg.~\ref{alg:bottom_up} instead of the standard TC-based kernel.

\begin{algorithm}[t]
\small
\caption{\acro{} Bottom-Up Kernel with CUDA Cores}
\label{alg:bottom_up}
\begin{algorithmic}[1]
\Require CSR ({\em Compressed Sparse Rows)}  of $G^\top$
\State \textbf{Stage 1: Lazy marking\ \hrulefill}
\For{\(w = \textit{warpID};\ w < \lceil n/32 \rceil;\ w \gets w + \#\text{warps}\)}
    \State $\textit{unvisitedMask} \gets \neg\, \mathbf{V}^{32}_{\text{curr}}[w]$
    \If{$(\textit{unvisitedMask} \gg \textit{laneID}) \mathbin{\&} 1$}
        \State $u \gets 32w + \textit{laneID}$
        \If{$u < n$}
            \For{$\textit{ptr} = \mathrm{rowPtrs}[u]$ \bf{to} $\mathrm{rowPtrs}[u+1] - 1$}
                \State $v \gets \mathrm{colIds}[\textit{ptr}]$
                \If{$\textbf{F}^1_{\text{curr}}[v] = 1$}
                    \State $\mathbf{V}^{32}_{\text{next}}[w] \stackrel{\text{atomic}}{\lor} (1 \ll \textit{laneID})$
                    \State \textbf{break}
                \EndIf
            \EndFor
        \EndIf
    \EndIf
\EndFor
\State \(\lvert \textbf{Q}_{\text{next}} \rvert \gets 0;\) \Comment{prepare for Stage 2}
\State \textsc{GridSync}();
\State \textbf{Stage 2: Frontier finalization} (same as Algorithm~\ref{alg:bvss_kernel_lazy})\ \hrulefill
\end{algorithmic}
\end{algorithm}


Finally, \acro relies on \textsc{GridSync} in  Algs.~\ref{alg:bvss_kernel},~\ref{alg:bvss_kernel_lazy}, and~\ref{alg:bottom_up}, which has been available via Cooperative Groups~\cite{cooperative_groups} and synchronizes the grid without returning control back to the host. This eliminates per-level kernel-launch overhead, whose performance impact is especially significant on graphs with thousands of BFS levels, such as road networks.

\section{\acro for Multi-Source BFS Workloads}
\label{sec:extension}

Although \acro can trivially perform concurrent BFSs with distinct sources, a careful profiling~\cite{ncu} of Alg.~\ref{alg:bvss_kernel_lazy} reveals that 39\% of the total execution time is spent waiting on accesses to BVSS. With an efficient multi-source execution, it is possible to overlap these memory accesses across concurrent BFSs. 

\subsection{Multi-Source BFS}
\label{subsec:msbfs}


Let $\kappa$ be the number of BFSs processed concurrently in a single kernel, and let $\rho$ be $\lceil n/\sigma \rceil$. In Alg.~\ref{alg:bvss_kernel_lazy}, the frontier and visited arrays encode a single BFS: bit $u$ is set if vertex $u$ is in the frontier or has been visited, respectively. The natural extension to $\kappa$ concurrent BFS instances stacks these arrays in structure-of-arrays (SoA) format, so that each $u$ is associated with a $\kappa$-bit element whose $c$-th bit encodes the frontier or visited status of $u$ in the $c$-th BFS. Bitwise operations such as $\textit{diff} = \mathbf{V}^{\kappa}_{\text{curr}}[u] \oplus \mathbf{V}^{\kappa}_{\text{next}}[u]$ then process all $\kappa$ BFS instances simultaneously in a single instruction, as in~\cite{sariyuce_kaya}. This naive adoption, however, introduces a memory coalescing problem.

In Alg.~\ref{alg:bvss_kernel_lazy}, each thread processes $4\sigma$ consecutive vertices (4 slice sets) and updates the corresponding frontier entries, with consecutive threads touching consecutive 4-byte words --- an access pattern that is fully coalesced. With $\kappa=256$, each $\kappa$-bit element occupies 32 bytes, so the stride between the elements touched by consecutive threads becomes $4\sigma\frac{\kappa}{8}$ bytes~($1$KB for $\sigma = 8$). We resolve this by (1) scheduling consecutive threads to consecutive slice sets (instead of words) and (2) introducing a bijective index remapping $\mathtt{getVI}(u,\rho) = (u \bmod \sigma) \cdot \rho + \lfloor u/\sigma \rfloor$, which contiguously stores the vertices with the same $u \bmod \sigma$. For example, with $n = 24$ and $\rho = 3$, the layout becomes $\underline{0, 8, 16}, \underline{1, 9, 17}, \ldots, \underline{7, 15, 23}$. Hence, the index $u \cdot \kappa + c$ is replaced by $\mathtt{getVI}(u,\rho) \cdot \kappa + c$. This allows consecutive threads to process consecutive slice sets, whose $\kappa$-bit elements are contiguous in memory, reducing the inter-thread stride to $\kappa/8$ bytes, i.e., 32 bytes for $\kappa=256$, and restoring coalesced accesses with only reindexing overhead.

When a slice set is active in at least one BFS, all of its VSSs are enqueued into the frontier. 
However, being active only for a single BFS
should not yield unnecessary lazy markings to be conducted for every other $\kappa - 1$ BFS instances. To address this, we introduce an array $\mathtt{activeSets}$, where $\mathtt{activeSets}[s]$ is a $\kappa$-bit element whose $c$-th bit is set iff slice set $s$ is active for the $c$-th BFS. 
During lazy marking, the warp iterates over the set bits of $\mathtt{activeSets}[s]$, and processes only the active BFSs. For efficiency, values from different BFSs are first accumulated into a thread-private buffer, and the resulting lazy atomic updates to $\mathbf{V}_{\text{next}}$ are issued to L2 cache only after all active BFSs have been processed. 
Hence, neither the lazy marking nor the frontier finalization introduces additional overhead. The algorithm is given in Alg.~\ref{alg:msbfs_kernel}. 

\algrenewcommand\algorithmicindent{1.1em}%
\begin{algorithm}[t]
\small
\caption{\acro{} (Generic) Multi-Source BFS Kernel}
\label{alg:msbfs_kernel}
\begin{algorithmic}[1]
\Require (1) BVSS data structure, (2) srcs: $\kappa$ source vertices $\{src_0, \ldots, src_{\kappa - 1}\}$
\State \(\mathtt{init}(\textbf{srcs}, \textbf{F}_{\text{curr}}, \textbf{V}_{\text{curr}}, \textbf{V}_{\text{next}}, \textbf{Q}_{\text{curr}}, \textbf{activeSets}, \textbf{dirtySets})\);
\State $\hat{n} \gets \lceil n/\sigma \rceil \cdot \sigma$; \Comment{$\sigma$-padded vertex count}
\While{\(\textit{cont}\)}
  \State \(\ell \gets \ell + 1;\)
  \State \textbf{Stage 1: Lazy marking\ \hrulefill}
  \For{$w = \textit{warpID};\ w < |\mathbf{Q}_{\text{curr}}|;\ w \gets w + \#\text{warps}$}
  
    \State \hspace*{0.4ex}... \\ \hspace*{5.2ex}Lines 11--15 of Algorithm~\ref{alg:bvss_kernel} \\ \hspace*{5.2ex}...
    \setcounter{ALG@line}{11}
    
    \IIf{$\textit{laneID} < \sigma$}
      $\textit{fv} \gets \mathbf{F}^{\kappa}_{\text{curr}}[\textit{laneID} \cdot \rho + ss_{in}]$;
    \EndIIf
    \IIf{$\textit{laneID} \geq \sigma$} $\textit{fv} \gets \mathbf{0}$; \EndIIf
    \State $\mathcal{B} \gets \mathbf{activeSets}[ss_{in}]$; \Comment{$\kappa$-bit active BFS mask}
    \State $\mathcal{M}[0..3] \gets \mathbf{0}$; \Comment{per-row register accumulation buffers}
    \While{$\mathcal{B} \neq 0$} \Comment{two TC calls per active BFS $c$}
      \State $c \gets \textbf{\_ffs}(\mathcal{B}) - 1$; \Comment{in $\{0,\dots,\kappa-1\}$}
      \State $\mathcal{B} \gets \mathcal{B} \mathbin{\&} (\mathcal{B} - 1)$; \Comment{unset $c$}
      \State $\textit{inc} \gets (\textit{fv} \gg c) \mathbin{\&} 1$;
      \State $\alpha \gets \texttt{warp.ballot}(\textit{inc})$;
      \Comment{reconstruct frontier}
      
      \State \hspace*{0.4ex}... \\ \hspace*{7.5ex}Lines 17--23 of Algorithm~\ref{alg:bvss_kernel} \\ \hspace*{7.5ex}...
      \setcounter{ALG@line}{27}
      
      \For{$q \in \{0,1,2,3\}$} \Comment{column in VSS matrix}
        \If{$\textit{fragC}[q] \neq 0$}
          \State $\mathcal{M}[q] \gets \mathcal{M}[q] \lor (1 \ll c)$; \Comment{accumulate}
        \EndIf
      \EndFor
    \EndWhile
    \For{$q \in \{0,1,2,3\}$} \Comment{flush register buffers to $\mathbf{V}_{\text{next}}$}
      \If{$\mathcal{M}[q] \neq \mathbf{0}$}
        \State $u \gets (u_0,u_1,u_2,u_3)[q]$;
        \State $\mathbf{V}^{\kappa}_{\text{next}}[\mathtt{getVI}(u,\rho)] \stackrel{\text{atomic}}{\lor} \mathcal{M}[q]$;
        \State $\mathbf{dirtySets}[u / \sigma] \gets \mathrm{1}$;
      \EndIf
    \EndFor
  \EndFor
  \State \(\lvert \textbf{Q}_{\text{next}} \rvert \gets 0;\) \Comment{prepare for Stage 2}
  \State \textsc{GridSync}();
  \State \textbf{Stage 2: Frontier Finalization\ \hrulefill}
  \For{$u_b = \textit{threadID} \cdot \sigma;\ u_b < \hat{n};\ u_b \gets u_b + \#\text{threads} \cdot \sigma$}
    \State $ss_{out} \gets u_b / \sigma$
    \If{$\mathbf{dirtySets}[ss_{out}] = \mathrm{0}$} \textbf{continue} \EndIf
    \State $\mathbf{dirtySets}[ss_{out}] \gets \mathrm{0}$;
    \State $\mathcal{A} \gets \mathbf{0}$;
    \For{$u = u_b, \ldots, u_b + \sigma - 1$}
      \State $\textit{idx} \gets \mathtt{getVI}(u, \rho)$;
      \State \(\textit{next} \gets {\textnormal{\textbf{V}}}^{\kappa}_{\text{next}}[idx];\)
      \State \(\textit{diff} \gets {\textnormal{\textbf{V}}}^{\kappa}_{\text{curr}}[idx] \,\mathbin{\oplus}\, \textit{next};\)\vspace*{0.7ex}
      \State $\mathbf{F}^\kappa_{\text{curr}}[\textit{idx}] \gets \textit{diff}$;
      \If{$\textit{diff} \neq \mathbf{0}$}
        \State $\mathbf{V}^{\kappa}_{curr}[\textit{idx}] \gets \textit{next}$;
        \State $\mathcal{A} \gets \mathcal{A} \lor \textit{diff}$;
      \EndIf
    \EndFor
    \State $\mathbf{activeSets}[ss_{out}] \gets \mathcal{A}$;
    \If{$\mathcal{A} \neq \mathbf{0}$}
      \State $[s,e) \gets [\mathbf{realPtrs}[ss_{out}],\ \mathbf{realPtrs}[ss_{out}+1])$;
      \State $\mathbf{Q}_{\text{next}} \stackrel{\text{warp-atomic}}{\gets} \mathbf{Q}_{\text{next}} \cup [s,e)$;
    \EndIf
  \EndFor
  
  \State ... \\ \hspace*{2.3ex}Lines 51--54 of Algorithm~\ref{alg:bvss_kernel_lazy}  
  
\EndWhile
\end{algorithmic}
\end{algorithm}

 Without reindexing, $\mathbf{F}^{\sigma}_{\text{curr}}[ss_{in}]$ is a contiguous $\sigma$-bit word storing the active/frontier status of the columns of slice set $ss_{in}$. However, with it, these $\sigma$ bits are no longer contiguously stored. To reconstruct them efficiently for the TC multiplication, the first $\sigma$ lanes of the warp gather $\mathbf{F}^{\kappa}_{\text{curr}}[\textit{laneID} \cdot \rho + ss_{in}]$~(line 12 of Alg.~\ref{alg:msbfs_kernel}). 
 The remaining $32 - \sigma$ lanes do not issue any memory access and initialize their buffer to zero (line 13). Each value fetched by the first $\sigma$ lanes is a $\kappa$-bit word encoding the full multi-source BFS frontier status, with bit $c$ of lane $u$'s word set if vertex $ss_{in} \cdot \sigma + u$ is in the frontier for BFS $c$. In each iteration over an active BFS $c$ (line 16), the warp performs a partial warp-level transpose via \texttt{warp.ballot}: each of the $\sigma$ lanes extracts bit $c$ from its local $\kappa$-bit buffer (line 19), contributing a single frontier bit to the collective ballot. Since lane $u \in \{0, \ldots, \sigma-1\}$ holds the frontier status of vertex $ss_{in} \cdot \sigma + u$ for all $\kappa$ BFSs, its vote corresponds to whether vertex $ss_{in} \cdot \sigma + u$ is in the frontier for BFS $c$. The result of \texttt{warp.ballot}, the $\sigma$-bit word $\alpha$, assembling these $\sigma$ votes, is broadcast to all lanes~(line 20). Then, the TC multiplication runs as in the single-source BFS.

The lazy vertex update scheme is a bottleneck for networks that are not scale-free-like; the per-level frontier is small and full swap in Stage~2 is expensive. To mitigate, we introduce \emph{dirty slice sets}; a slice set is marked {\em{dirty}} during Stage~1 whenever any of its vertices is lazily marked in $\mathbf{V}_{\text{next}}$, signaling that it requires processing in Stage~2. Dirty slice sets are built during lazy marking and consumed during frontier finalization; if a slice set is not dirty, it is skipped entirely in Stage~2, since none of its vertices were marked and no new frontier entries can reside within it. The dirty and active slice sets are duals: the former is produced in Stage~1 to accelerate Stage~2, while the latter is vice versa. 

Alg.~\ref{alg:msbfs_kernel} is kept generic to use for kernels such as closeness centrality. Hence, the level updates which are not necessary in those are not included. If they were, the statement $\textbf{level}$[$\kappa \times b + idx$] $= \ell$, which sets $u$'s level for the $b$th BFS--for every set bit of $diff$ at index $b$--would be within the {\bf if} at line 49. 

\subsection{Closeness Centrality}
\label{subsec:cc}

To compute closeness centrality of all vertices, one can run an {\em all-pairs shortest-path computation} on $G$. When $G$ is connected, undirected, and unweighted, this is equivalent to performing BFSs from all source vertices, which requires an efficient multi-source BFS kernel. Within the {\bf{if}} statement at line 49, the required values can be captured as
\begin{equation}
    \mathbf{far}[idx] \gets \mathbf{far}[idx] + \ell \cdot \mathrm{popc}(\textit{diff}).
    \label{eq:far}
\end{equation}
After accumulating the $\mathbf{far}$ values across all BFSs, the centrality can be computed via:
\begin{equation}
    \mathrm{cc}[u] = (n-1) / \mathbf{far}[\mathtt{getVI}(u, \rho)].
    \label{eq:cc}
\end{equation}
Since Alg.~\ref{alg:msbfs_kernel} processes $\kappa$ concurrent BFS instances per launch, it is invoked $\lceil n/\kappa \rceil$ times, each time from a distinct batch of $\kappa$ sources, to perform all necessary distance computations. For a disconnected graph, a different normalization using component sizes can be applied to each centrality score.

\subsection{Using Tensor Cores for Other Graph Algorithms}
\label{sec:anatomy}

A binary TC multiplication provides a single bit of information for two (binary) vectors: whether or not {\em there exists an index at which both vectors simultaneously have a set bit}. For graph algorithms in which information transmission can be conducted entirely on this single bit, TCs are well-suited and promise performance improvements. BFS, Multi-Source BFS, and Closeness Centrality---as well as the ones not covered in this work, such as Triangle Counting---satisfy this property. BFS focuses on {\em the existence of two set bits at the same position of the frontier vector and the connectivity mask}, i.e., whether the pulling vertex has an active frontier neighbour. Since this transmission involves only one bit, BFS maps onto TCs.

Not every graph algorithm, however, depends on information that can be encoded in a single bit. Consider a hypothetical adaptation of Alg.~\ref{alg:bvss_kernel_lazy} to \emph{Connected Components}. Instead of $\mathbf{L}$, one can maintain a $\mathbf{comp}$ array, where $\mathbf{comp}[u] = \mathbf{comp}[v]$ if vertices $u$ and $v$ are connected. A trivial, synchronized implementation initially enqueues all VSSs in the frontier and sets $\mathbf{comp}[u] = u$, $u \in \mathcal{V}$. Whenever a TC produces a nonzero result at line~27 of Alg.~\ref{alg:bvss_kernel_lazy}, it signals a state change for $u$; $\mathbf{comp}[u] \leftarrow \mathbf{comp}[v]$ for {\em some} column/vertex $v$ of the VSS. However, although a {\em state change signal} is a one-bit value, one must also identify $v$ to retrieve the {\em actual information} $\mathbf{comp}[v]$ to be transmitted into $\mathbf{comp}[u]$. This necessitates an explicit bitwise traversal of the $\sigma$-bit slice to locate the column $v$, which makes the kernel inefficient. On the contrary, for BFS and closeness centrality, the information to be transmitted, $\mathbf{L}[v]$, i.e., the distance to the BFS source, is the same regardless of the VSS column $v$. 

A similar argument can also be made for kernels whose mechanics are more similar to BFS, such as {\em Betweenness Centrality}. In a Brandes-like implementation~\cite{Brandes01062001}, the state change during the first stage implies $\mathbf{npaths}[u]$ += $\mathbf{npaths}[v]$ where $\mathbf{npaths}[u]$ is the number of shortest paths from the BFS source to $u$ for all $u \in \mathcal{V}$. To correctly perform this update, one needs to locate $v$. This is also true for BFS implementations that lack level synchronization, for which the active frontier queue can accommodate vertices belonging to different BFS levels. For all these graph kernels, the techniques used by \acro need to be extended as they cannot be applied as efficiently as they currently are. 



\section{Experiments}
\label{sec:experiments}

\acro\footnote{\url{https://github.com/delbek/blest}} is implemented in C++/CUDA and compiled with \texttt{gcc~12.3.0} and \texttt{CUDA~13.0}. Experiments are conducted on two servers, \textbf{Arch-1} and \textbf{Arch-2}. \textbf{Arch-1} is equipped with a 64-core Intel Xeon Gold 6548Y+ (2.5~GHz) and 1~TB of host RAM, paired with an NVIDIA H200 GPU (141~GB HBM3e) on the device side. \textbf{Arch-2} is equipped with two Intel Xeon Platinum 8460Y+ processors (40 cores each, 2.0~GHz) and 512~GB of host RAM, paired with NVIDIA H100 GPUs (64~GB HBM2e) on the device side. All experiments are conducted on \textbf{Arch-1}, with the exception of the Closeness Centrality experiment (Table~\ref{tab:closeness_centrality}), which is conducted on \textbf{Arch-2}.\looseness=-1

We use two benchmark suites: (1) the GAP Benchmark Suite~\cite{gap}, and (2) a custom benchmark suite containing large graphs from SuiteSparse~\cite{suitesparse}, comprising \underline{all} graphs satisfying $|\mathcal{V}| \geq 23M$ and $|\mathcal{E}| \leq 2^{32} - 1$. When multiple graphs from the same matrix group meet these criteria, only the largest is retained. Graphs already included in the GAP suite are deduplicated, yielding 14 graphs in total.

Table~\ref{tab:ssbfs_blest_vs_sota} presents the single-source BFS (SS-BFS) performance of \acro against SotA implementations. \acro decisively outperforms all baselines across all graphs, with average speedups of $22.0\times$, $7.7\times$, $8.1\times$, and $5.9\times$ over GAP~\cite{gap}, Gunrock~\cite{gunrock}, GSWITCH~\cite{gswitch}, and BerryBees~\cite{berrybees}, respectively. Even \cite{berrybees}---the only other TC-based SS-BFS kernel and the fastest prior implementation in the literature---is on average $5.9\times$ slower than \acro, establishing \acro as the new state-of-the-art in the SS-BFS literature.

\begin{table}[h]
\centering
\caption{\small
Performance (in ms.) of \acro against SotA SS-BFS implementations. All runtimes are averaged over 64 random sources. All the speedups are normalized to \cite{berrybees}. Some baselines failed on  large graphs due to unsupported graph size or out-of-memory issues. These cases are marked with (err.) in the table. 
}
\scalebox{0.78}{
\setlength{\tabcolsep}{1.6pt}
\begin{tabular}{c|lc||rr||rrrr||r|r}
\multicolumn{5}{c}{} & \multicolumn{6}{c}{\textbf{Times (in milliseconds)}} \\ \cline{6-11}
& \textbf{Graph} & \multicolumn{1}{c||}{
\rotatebox{90}{\parbox{1.6cm}{\centering \textbf{Scale-Free}}}
} & \textbf{$|\mathcal{V}|$} & \textbf{$|\mathcal{E}|$} &
\multicolumn{1}{c}{
\rotatebox{90}{\parbox{1.6cm}{\centering \textbf{GAP}\\\cite{gap}}}
} &
\multicolumn{1}{c}{
\rotatebox{90}{\parbox{1.6cm}{\centering \textbf{Gunrock}\\\cite{gunrock}}}
} &
\multicolumn{1}{c}{
\rotatebox{90}{\parbox{1.8cm}{\centering \textbf{GSWITCH}\\\cite{gswitch}}}
} &
\multicolumn{1}{c||}{
\rotatebox{90}{\parbox{1.6cm}{\centering \textbf{BerryBees}\\\cite{berrybees}}}
} &
\multicolumn{1}{c|}{
\rotatebox{90}{\parbox{1.6cm}{\centering \textbf{\acro}\\(this work)}}
} &
\rotatebox{90}{\parbox{1.6cm}{\centering \textbf{\acro}\\vs.~\cite{berrybees}}} \\
\midrule
\multirow{5}{*}{\rotatebox[origin=c]{90}{GAP}}
&{\tt{GAP-road}}    & \text{\sffamily X} & 23M  & 57M  & 739.6  & 317.5  & 130.2  & 161.9  & 41.3  & 3.9$\times$ \\
&{\tt{GAP-twitter}} & $\checkmark$       & 61M  & 1.4B & 201.5  & (err.) & (err.) & 63.7   & 6.6   & 9.7$\times$ \\
&{\tt{GAP-web}}     & $\checkmark$       & 50M  & 1.9B & 327.1  & (err.) & 277.8  & 74.3   & 15.7  & 4.7$\times$ \\
&{\tt{GAP-kron}}    & $\checkmark$       & 134M & 4.2B & 290.2  & (err.) & (err.) & 164.9  & 7.2   & 23.1$\times$ \\
&{\tt{GAP-urand}}   & $\checkmark$       & 134M & 4.2B & 468.8  & (err.) & (err.) & 192.9  & 13.4  & 14.4$\times$ \\
\midrule
\multirow{9}{*}{\rotatebox[origin=c]{90}{$|\mathcal{V}| \ge 23M, |\mathcal{E}| \le 2^{32} - 1$}}
&{\tt{nlpkkt240}}     & $\checkmark$       & 27M  & 760M & 234.3  & 25.1   & 66.4   & 33.3   & 8.9   & 3.7$\times$ \\
&{\tt{uk-2005}}       & $\checkmark$       & 39M  & 936M & 177.2  & 331.8  & 36.3   & 22.8   & 8.8   & 2.6$\times$ \\
&{\tt{it-2004}}       & $\checkmark$       & 41M  & 1.1B & 216.2  & (err.) & 97.1   & 21.4   & 10.1  & 2.1$\times$ \\
&{\tt{europe\_osm}}   & \text{\sffamily X} & 50M  & 108M & 1573.6 & 942.1  & 417.2  & 481.9  & 137.5 & 3.5$\times$ \\
&{\tt{com-Friends.}}  & $\checkmark$       & 65M  & 3.6B & 347.0  & (err.) & (err.) & 135.6  & 11.5  & 11.8$\times$ \\
&{\tt{Spiel.\_k600}}  & \text{\sffamily X} & 72M  & 216M & 520.9  & 40.5   & 231.2  & 181.3  & 19.4  & 9.3$\times$ \\
&{\tt{webbase-2001}}  & $\checkmark$       & 118M & 1.0B & 292.0  & 72.8   & 74.7   & 29.2   & 12.1  & 2.4$\times$ \\
&{\tt{kmer\_V1r}}     & \text{\sffamily X} & 214M & 465M & 1493.7 & 60.0   & 244.6  & 317.9  & 45.3  & 7.0$\times$ \\
&{\tt{mawi}}          & $\checkmark$       & 226M & 480M & 272.9  & 882.9  & 107.1  & 457.3  & 55.5  & 8.3$\times$ \\
\midrule\midrule
\multicolumn{3}{c}{Speedups are given}    & \multicolumn{2}{|c||}{Min}     & 0.1$\times$ & 0.1$\times$ & 0.2$\times$ & 1$\times$ & \multicolumn{2}{r}{2.1$\times$} \\
\multicolumn{3}{c}{over the graphs the}   & \multicolumn{2}{|c||}{Max}     & 1.7$\times$ & 5.3$\times$ & 4.3$\times$ & 1$\times$ & \multicolumn{2}{r}{23.1$\times$} \\
\multicolumn{3}{c}{kernels could process.}& \multicolumn{2}{|c||}{Geomean} & 0.3$\times$ & 0.8$\times$ & 0.7$\times$ & 1$\times$ & \multicolumn{2}{r}{5.9$\times$}
\end{tabular}}
\label{tab:ssbfs_blest_vs_sota}
\end{table}

For the Multi-Source BFS (MS-BFS), the most recent GPU-based implementation is not compatible with modern GPUs~\cite{ibfs}. 
Although CPU-based, another well-known implementation~\cite{the_more_the_merrier} delivers good performance on 64 cores of \textbf{Arch-1}, yet it is $25.9\times$ slower on average than \acro for the graphs in Table~\ref{tab:ssbfs_blest_vs_sota}. Given the current state of the MS-BFS literature, \acro is all the more valuable: it is engineered for modern GPUs and delivers performance that is $2.7\times$ faster on average than its own single-source variant. 

Since computing closeness centrality for all the vertices has $\mathcal{O}$($nm$) complexity,   approximation algorithms~\cite{approx1, approx2} that apply sampling, pivoting, pruning, or top-$k$ approaches are used at the cost of accuracy. With \acro, it is now possible to compute \underline{exact} values for all vertices in real-world graphs. To demonstrate \acro's MS-BFS performance, we compute exact scores for {\tt com-Friendster} from Table~\ref{tab:ssbfs_blest_vs_sota}, which is, to the best of our knowledge, the largest graph for which exact centrality values have ever been computed. With 100 H100s, computing the closeness centrality of {\tt com-Friendster} took 3{,}665 seconds, i.e., roughly 100 GPU-hours. For this experiment, sources are partitioned via MPI into disjoint batches across GPUs and the partial sums are reduced in a final stage. We additionally compute the closeness centrality values for five other graphs---two social, one geometric, one random, and one road network---given in Table~\ref{tab:closeness_centrality}.

\begin{table}[h]
\centering
\caption{\small Closeness centrality performance of \acro across six graphs. Execution times are reported in seconds.}
\scalebox{0.73}{
\setlength{\tabcolsep}{2pt}
\begin{tabular}{r|lrrr|lrrr}
\textbf{GPUs} & \textbf{Graph} & \textbf{$|\mathcal{V}|$} & \textbf{$|\mathcal{E}|$} & \textbf{Time}  & \textbf{Graph} & \textbf{$|\mathcal{V}|$} & \textbf{$|\mathcal{E}|$} & \textbf{Time} \\
\midrule
5   & {\tt com-LiveJ.}    & 4.0M  & 69.4M  & 118     & {\tt com-Orkut}         & 3.1M  & 234.4M & 131      \\
10  & {\tt delaun.\_n24}      & 16.8M & 100.7M & 4{,}836  & {\tt rgg\_n\_2\_24} & 16.8M & 265.1M & 4{,}711 \\
100 & {\tt com-Friend.}     & 65.6M & 3.6B   & 3{,}665  & {\tt GAP-road}          & 23.9M & 57.7M  & 1{,}701 \\
\end{tabular}}
\label{tab:closeness_centrality}
\end{table}

\subsection{Ablation Studies}
\label{subsec:ablation}

We evaluate the benefits of the proposed optimizations with an ablation study for both SS-BFS and MS-BFS variants. For SS-BFS, we construct five variants of \acro, starting from the base \acro~(A) and incrementally adding one optimization at a time until the final version is reached. This ablation study is summarized in Table~\ref{tab:ssbfs_ablation}. \acro~(A) is the initial variant, incorporating only the BVSS data structure and kernel fusion, which achieves a $1.6\times$ average speedup over \cite{berrybees}~(col. 3). As optimizations are added one by one, the speedup increases, with the final version of \acro achieving an average $5.9\times$ speedup over \cite{berrybees} across all the graphs. The ablation study clearly demonstrates that each optimization may contribute, although the impact varies across graphs. For instance, reordering is effective on 8/14 graphs, neutral on 2/14, and degrades performance on 4/14.

\begin{table}[h]
\centering
\caption{\small SS-BFS ablation study of \acro (in ms., averaged over 64 random sources). (A): BVSS + kernel fusion. (AB): (A) + optimal TC layout. (ABC): (AB) + reordering. (ABCD): (ABC) + lazy vertex updates. (Full): (ABCD) + switching. N/A indicates the optimization is inapplicable. For each graph/column, the speedup w.r.t.~\cite{berrybees} is computed using the latest non-N/A runtime among all variants up to and including that column.}
\scalebox{0.75}{
\setlength{\tabcolsep}{1.5pt}
\begin{tabular}{c|l|c||r||rrrrr}
\multicolumn{3}{c}{}& \multicolumn{6}{c}{\textbf{Times (in milliseconds)}} \\ \cline{4-9}
& & \multicolumn{1}{c||}{\textbf{Pseu.}}  &
\multicolumn{1}{c||}{\textbf{Berry}} &
\multicolumn{5}{c}{\textbf{\acro}} \\
& \textbf{Graph} &
\multicolumn{1}{c||}{\textbf{Diam.}} &
\textbf{Bees}~\cite{berrybees} &
\multicolumn{1}{c}{\textbf{(A)}} &
\multicolumn{1}{c}{\textbf{(AB)}} &
\multicolumn{1}{c}{\textbf{(ABC)}} &
\multicolumn{1}{c}{\textbf{(ABCD)}} &
\textbf{(Full)} \\
\midrule
\multirow{5}{*}{\rotatebox[origin=c]{90}{GAP}}
&{\tt{GAP-road}}      & 8.4K  & 161.9 & 62.8  & 49.3  & 41.3  & N/A  & N/A  \\
&{\tt{GAP-twitter}}   & 16    & 63.7  & 66.2  & 54.9  & 41.3  & 18.8 & 6.6  \\
&{\tt{GAP-web}}       & 64    & 74.3  & 19.1  & 17.0  & 16.7  & 15.7 & N/A  \\
&{\tt{GAP-kron}}      & 8     & 164.9 & 177.7 & 154.4 & 74.3  & 49.6 & 7.2  \\
&{\tt{GAP-urand}}     & 7     & 192.9 & 236.0 & 202.6 & 201.3 & 52.3 & 13.4 \\
\midrule
\multirow{9}{*}{\rotatebox[origin=c]{90}{$|\mathcal{V}| \ge 23M$, $|\mathcal{E}| \le 2^{32} - 1$}}
&{\tt{nlpkkt240}}     & 242   & 33.3  & 14.9  & 11.2  & 9.0   & 8.9  & N/A  \\
&{\tt{uk-2005}}       & 210   & 22.8  & 13.0  & 11.7  & 14.2  & 8.8  & N/A  \\
&{\tt{it-2004}}       & 65    & 21.4  & 14.8  & 13.3  & 14.4  & 10.1 & N/A  \\
&{\tt{europe\_osm}}   & 30.1K & 481.9 & 198.8 & 160.4 & 137.5 & N/A  & N/A  \\
&{\tt{com-Friendster}}& 26    & 135.6 & 173.3 & 146.2 & 107.5 & 53.5 & 11.5 \\
&{\tt{Spielman\_k600}}& 601   & 181.3 & 138.0 & 130.7 & 14.5  & 19.4 & N/A  \\
&{\tt{webbase-2001}}  & 632   & 29.2  & 27.2  & 25.3  & 30.6  & 12.1 & N/A  \\
&{\tt{kmer\_V1r}}     & 589   & 317.9 & 370.9 & 366.3 & 168.4 & 45.3 & N/A  \\
&{\tt{mawi}}          & 7     & 457.3 & 56.7  & 42.6  & 57.2  & 55.5 & N/A  \\
\midrule\midrule
  & \multicolumn{2}{c||}{Min Speedup}     & 1.0$\times$ & 0.8$\times$ & 0.9$\times$ & 1.0$\times$ & 2.1$\times$ & 2.1$\times$  \\
  & \multicolumn{2}{c||}{Max Speedup}     & 1.0$\times$ & 8.1$\times$ & 10.7$\times$ & 12.5$\times$ & 9.3$\times$ & 22.9$\times$ \\
  & \multicolumn{2}{c||}{Geomean}         & 1.0$\times$ & 1.6$\times$ & 1.9$\times$ & 2.5$\times$ & 3.9$\times$ & 5.9$\times$  \\
  \end{tabular}}
\label{tab:ssbfs_ablation}
\end{table}

Some optimizations are marked N/A for certain graphs, as they are designed for graphs meeting specific conditions. Direction switching is inapplicable to graphs whose per-level frontier never satisfies Eq.~\eqref{eq:switching}, and the corresponding~(\textbf{Full}) column entry is marked N/A. Similarly, the lazy vertex update scheme~(Sec.~\ref{subsec:lazy}) is developed specifically for scale-free-like networks whose update divergence is high and cannot be reduced via reordering algorithms such as RCM due to their inherently high bandwidth. For graphs that do not require the lazy scheme, Alg.~\ref{alg:bvss_kernel} is used exclusively, and the corresponding ABCD entry in Tab.~\ref{tab:ssbfs_ablation} is marked N/A. The dispatch decision is made based on $\mathcal{U}_{\mathrm{div}}$: graphs with $\mathcal{U}_{\mathrm{div}} > 25{,}000$ are dispatched to the lazy kernel (Alg.~\ref{alg:bvss_kernel_lazy}), while those with $\mathcal{U}_{\mathrm{div}} \leq 25{,}000$ are dispatched to Alg.~\ref{alg:bvss_kernel}.

The difference between (A) and (AB) lies in the layout; in (A), the one in \cite{berrybees} is used, requiring 16 TC multiplications per 128 slices. (AB) replaces this by the layout proposed in Sec.~\ref{subsec:layout}, requiring only 2 TC multiplications. Both variants use TCs as the compute unit, though the latter utilizes them $8\times$ more efficiently. Although the average speedup is only $1.2\times$, it is consistent over all the graphs. 

\begin{figure}[h]
  \centering
  \includegraphics[width=\linewidth]{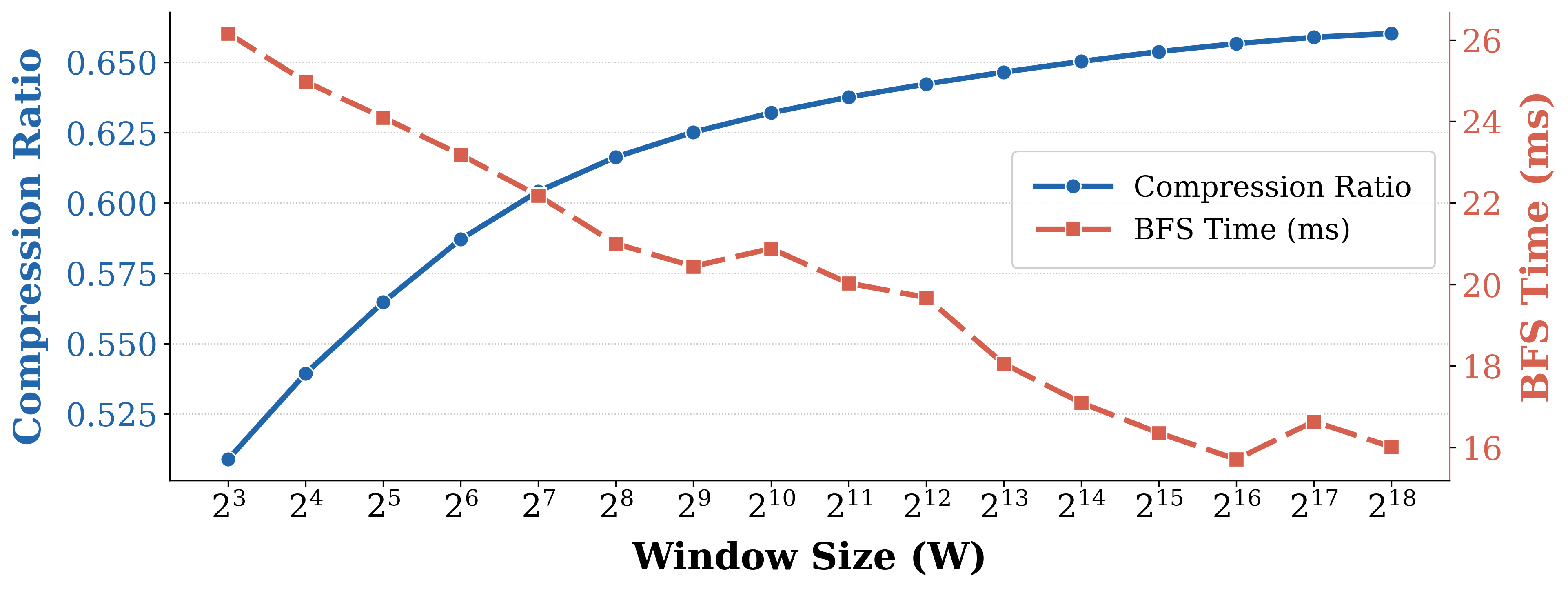}
  \caption{\small Effect of window size \(W\) on compression ratio and BFS runtime on {\tt GAP-web}.}
  \label{fig:jaccard-window-compression-time}
\end{figure}

Although Tab.~\ref{tab:ssbfs_ablation} shows that reordering degrades performance on 4/14 graphs, this does not imply reordering being harmful, but rather that the graphs favor natural orderings that reordering cannot improve upon. To validate this, we randomly reorder 6 graphs---including 4/14 where reordering is harmful and the 2/14 where it is neutral---and measure the performance of (ABC) against the randomly reordered variant (AB) in Tab.~\ref{tab:random_vs_ordering}. Except for {\tt mawi} and {\tt GAP-urand}, reordering is effective, although its real benefit is limited due to good natural vertex orderings. Since the reordering requires a hyperparameter $W$ to be set, we evaluate its sensitivity on {\tt GAP-web} by varying $W \in [2^3, 2^{18}]$ and analyzing the compression ratio and BFS execution time in Figure~\ref{fig:jaccard-window-compression-time}. As Fig.~\ref{fig:jaccard-window-compression-time} shows, increasing $W$ has a positive impact on both the compression ratio and BFS performance, as expected. However, the relationship is clearly concave-down, signaling diminishing returns as $W$ grows. Since larger $W$ also increases the reordering cost, the choice of $W$ requires balancing compression quality against preprocessing overhead. In all our experiments, we set $W = 2^{16}$.

\begin{table}[h]
\centering
\setlength{\tabcolsep}{2pt}
\caption{\small
Effect of replacing the natural ordering by a random reordering on graphs where (ABC) degrades the performance over (AB). Runtimes are in milliseconds.
}
\scalebox{0.80}{
\begin{tabular}{lrr|lrr}
& \multicolumn{1}{l}{\textbf{Rnd. Order}} & \multicolumn{1}{l|}{\textbf{\acro{}}}  & & \multicolumn{1}{l}{\textbf{Rnd. Order}} & \multicolumn{1}{l}{\textbf{\acro{}}} \\
\textbf{Graph} & \multicolumn{1}{l}{\textbf{\acro{} (AB)}} & \multicolumn{1}{l|}{\textbf{(ABC)}} & \textbf{Graph} & \multicolumn{1}{l}{\textbf{\acro{} (AB)}} & \multicolumn{1}{l}{\textbf{(ABC)}} \\
\midrule
{\tt{GAP-web}}      & 108.77 & 16.66 &
{\tt{it-2004}}      &  75.82 & 14.42 \\
{\tt{webbase-2001}} & 107.21 & 30.63 &
{\tt{uk-2005}}      &  64.07 & 14.16 \\
{\tt{mawi}}         &  55.47 & 57.24 & 
{\tt{GAP-urand}}    &  201.83 & 201.30
\end{tabular}}
\label{tab:random_vs_ordering}
\end{table}

\begin{figure*}[!htbp]
    \centering

    \begin{subfigure}[b]{0.33\textwidth}
        \centering
        \includegraphics[width=\textwidth]{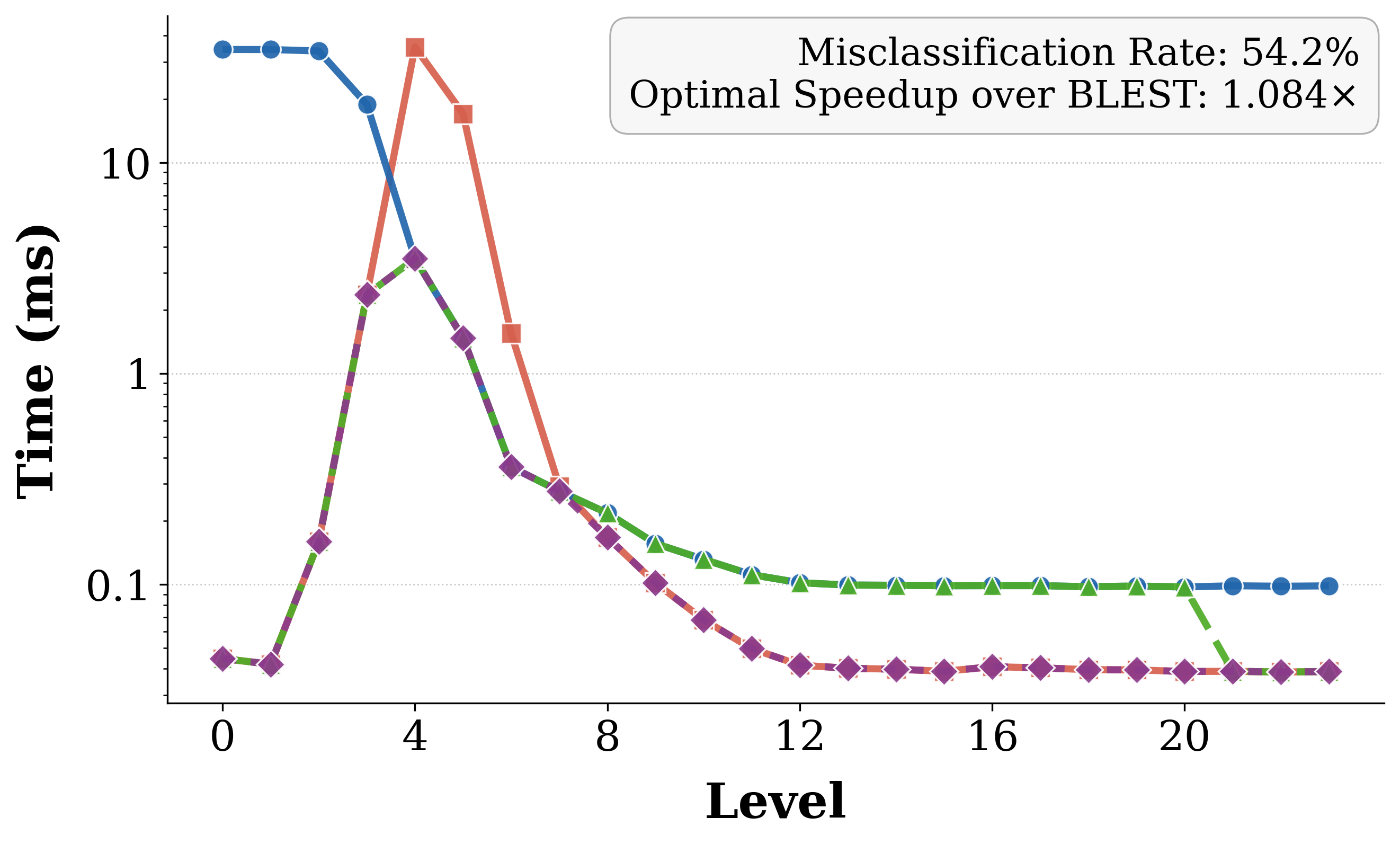}
        \caption{com-Friendster}
        \label{fig:switch:friendster}
    \end{subfigure}
    \hfill
    \begin{subfigure}[b]{0.33\textwidth}
        \centering
        \includegraphics[width=\textwidth]{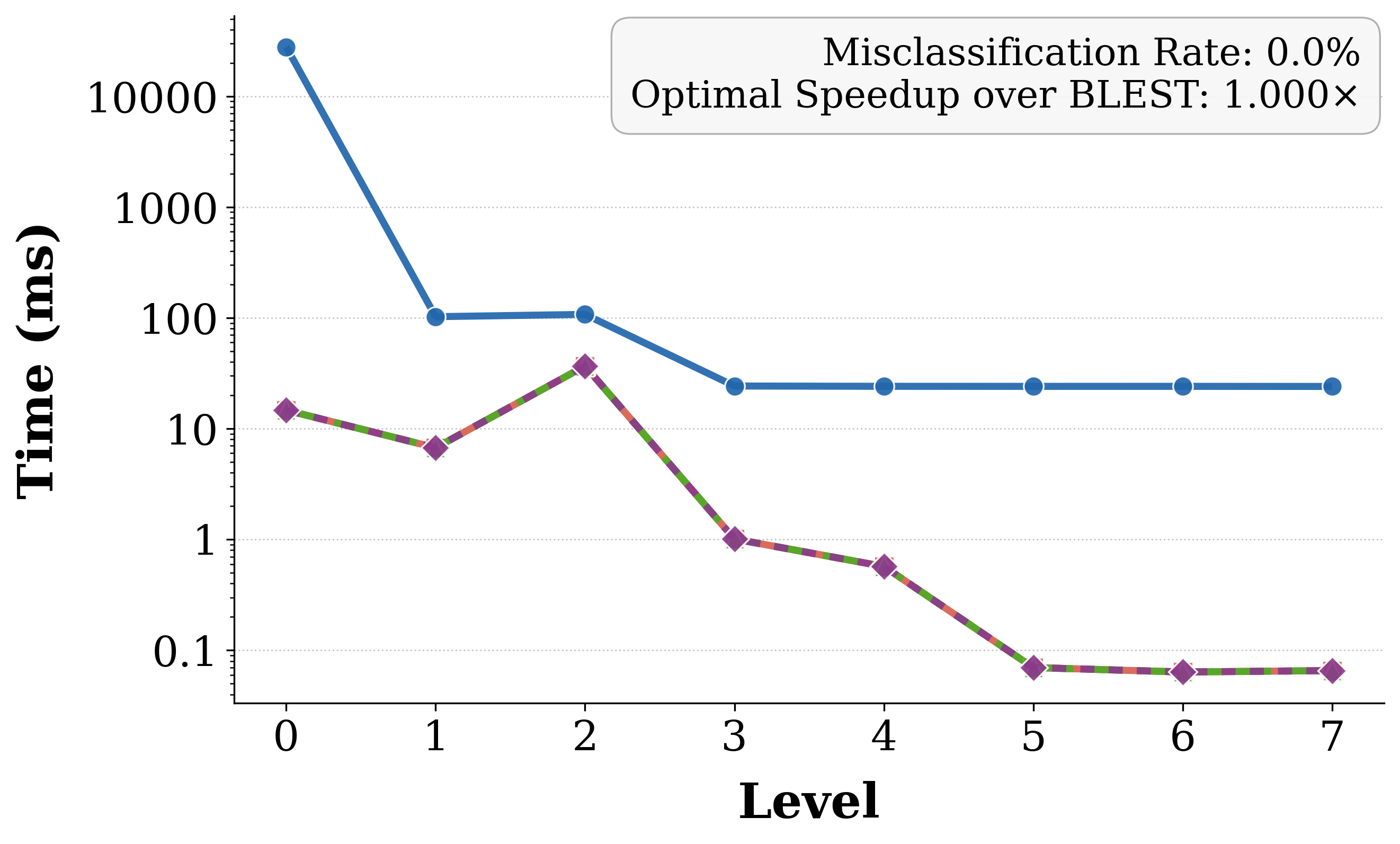}
        \caption{mawi}
        \label{fig:switch:mawi}
    \end{subfigure}
    \hfill
    \begin{subfigure}[b]{0.33\textwidth}
        \centering
        \includegraphics[width=\textwidth]{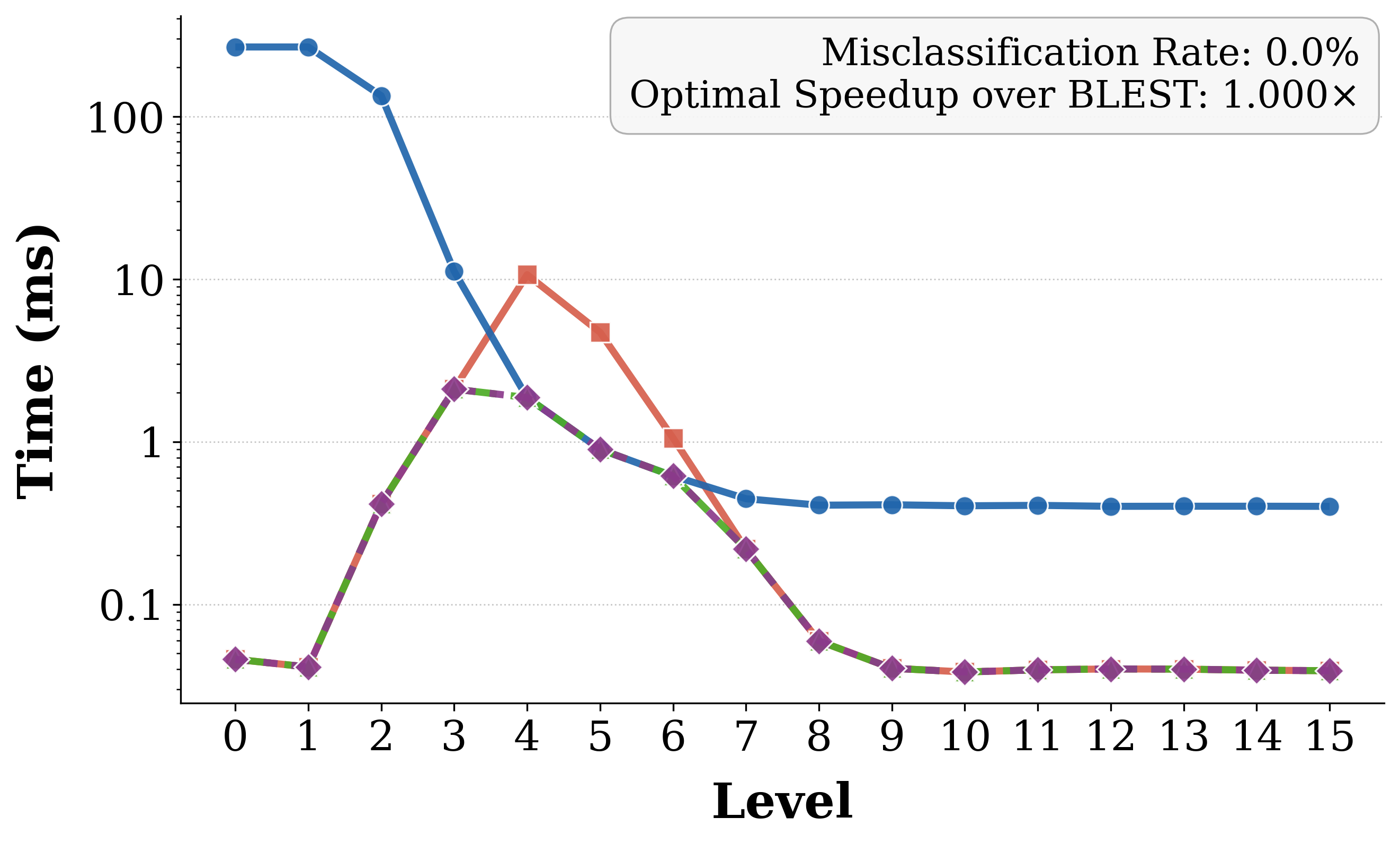}
        \caption{GAP-twitter}
        \label{fig:switch:twitter}
    \end{subfigure}


    \begin{subfigure}[b]{0.33\textwidth}
        \centering
        \includegraphics[width=\textwidth]{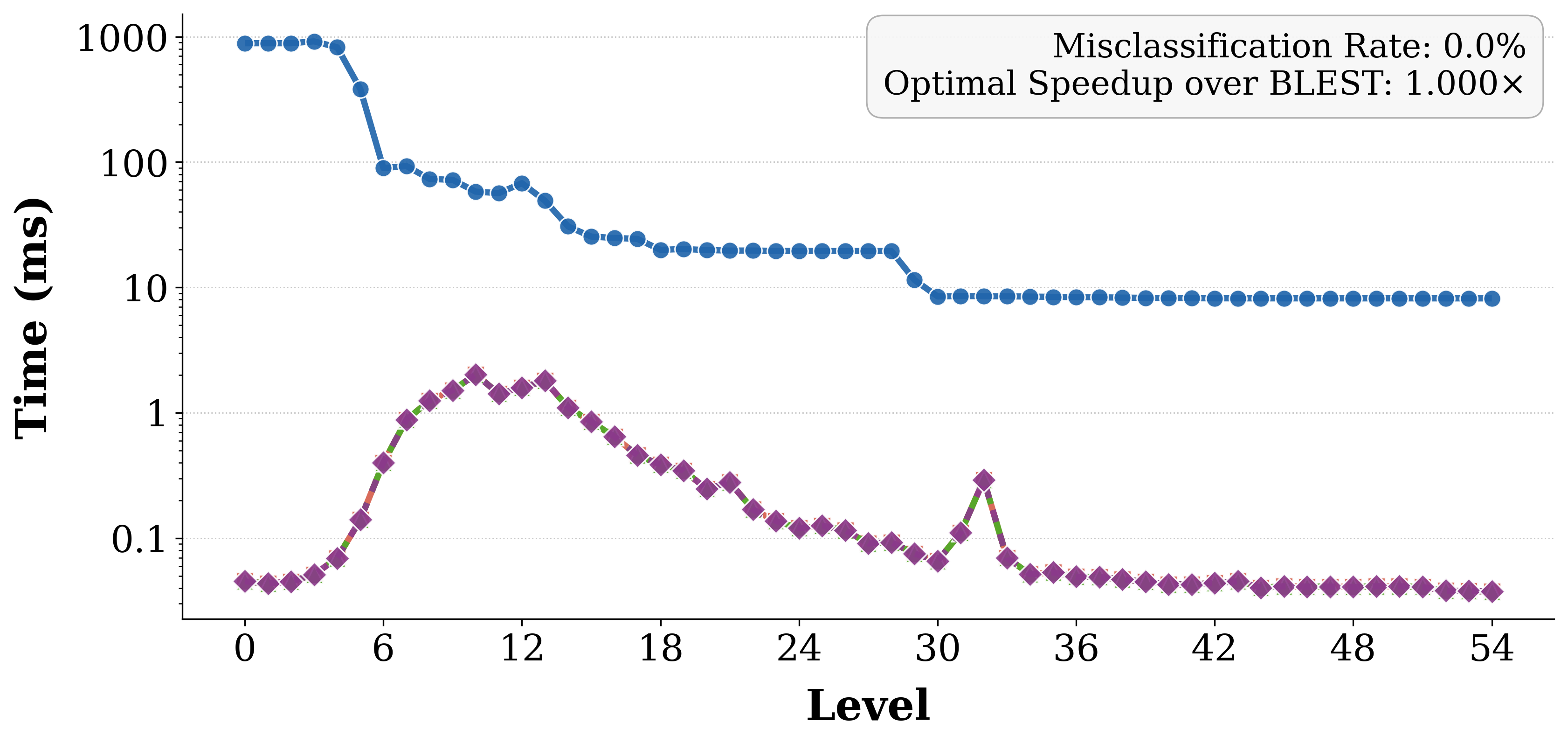}
        \caption{GAP-web}
        \label{fig:switch:web}
    \end{subfigure}
    \hfill
    \begin{subfigure}[b]{0.33\textwidth}
        \centering
        \includegraphics[width=\textwidth]{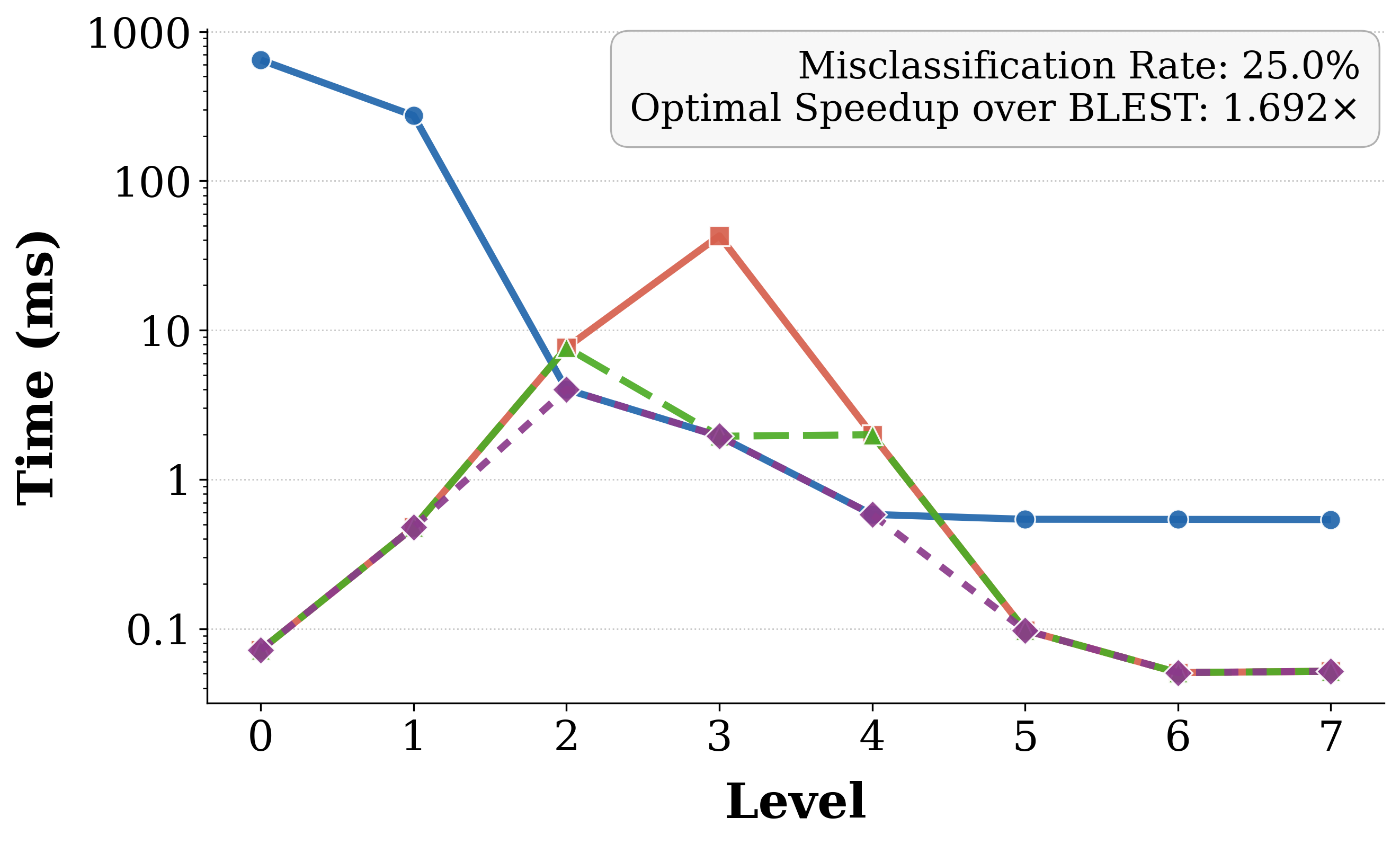}
        \caption{GAP-kron}
        \label{fig:switch:kron}
    \end{subfigure}
    \hfill
    \begin{subfigure}[b]{0.33\textwidth}
        \centering
        \includegraphics[width=\textwidth]{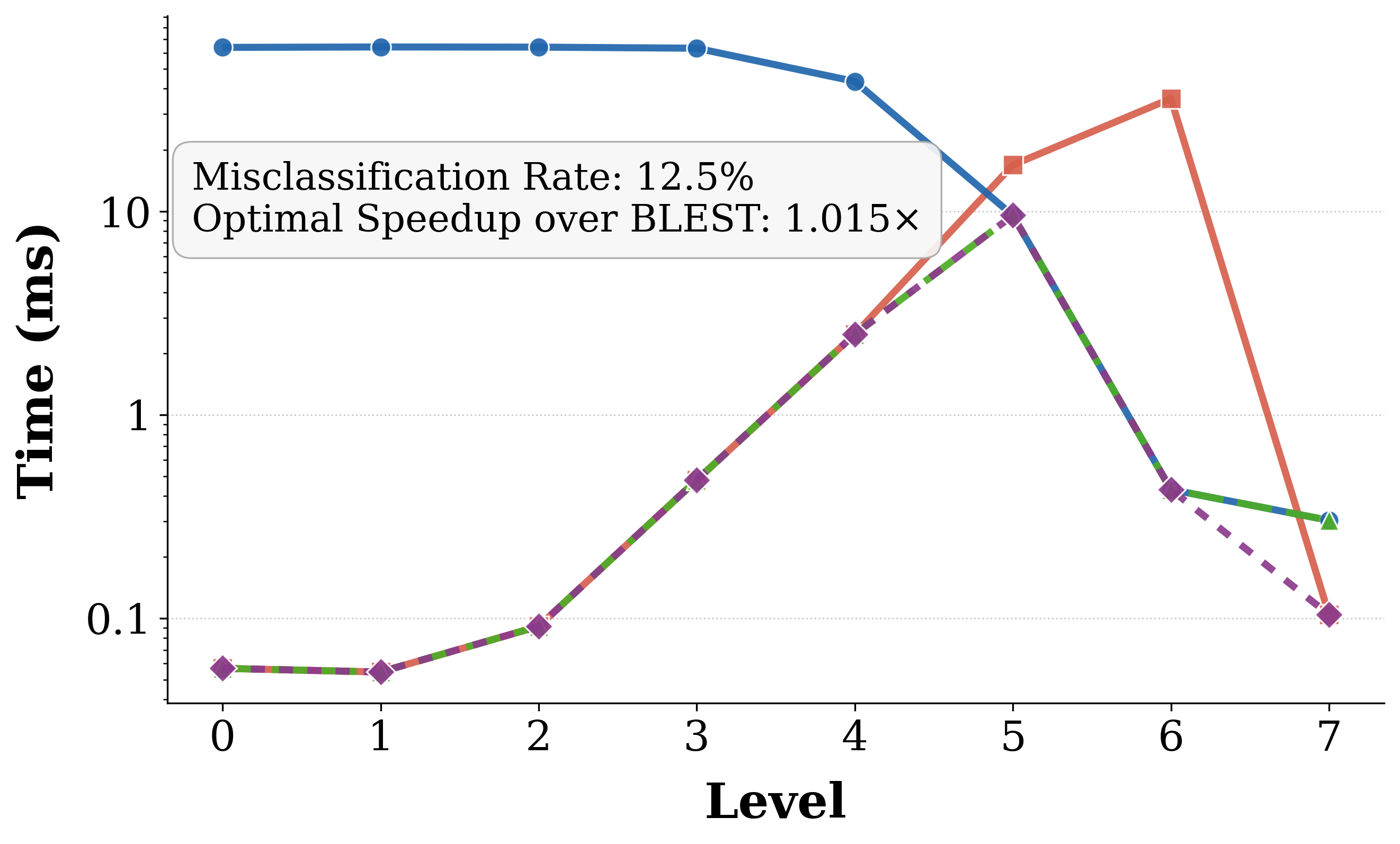}
        \caption{GAP-urand}
        \label{fig:switch:urand}
    \end{subfigure}
    
    \centering
    \includegraphics[width=0.5\textwidth]{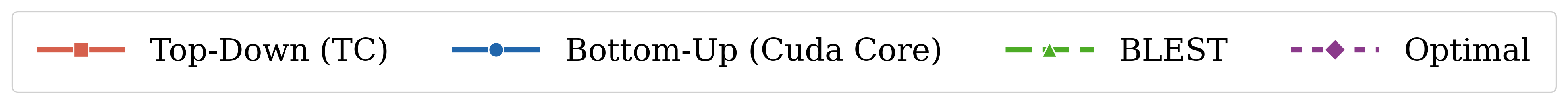}
    
    \caption{\small Switching analysis across 6 graphs. Each subplot shows the execution time (ms) of {\em Top-Down}, {\em Bottom-Up}, \acro, and {\em Optimal} for each BFS level. The {\em Optimal} curve is simply $\min(\text{Top-Down},\,\text{Bottom-Up})$ for each level, representing ideal switching. The misclassification rates and speedups indicate the fraction of levels at which \acro is wrong and the speedup of {\em Optimal} over \acro.}
    \label{fig:switching:all}
\end{figure*}

The impact of switching depends heavily on the per-level frontier size, as expressed in~\eqref{eq:switching}. Consequently, graphs with a small diameter are the most promising candidates for switching to the bottom-up CUDA Core implementation described in Section~\ref{subsec:switching}. Figure~\ref{fig:switching:all} analyzes the switching behavior of the 6 graphs from Tab.~\ref{tab:ssbfs_ablation} with the lowest pseudo-diameters, computed by finding 64 pseudo-peripheral vertices and executing BFS from each to identify the traversal with the greatest depth. The modes are: {\em Top-Down} (TC), {\em Bottom-Up} (CUDA Core), \acro (the policy determined automatically via \eqref{eq:switching}), and {\em Optimal} (a hypothetical oracle selecting the correct mode at each level). For graphs where \acro determines that switching is unnecessary,---by executing 3 BFS runs from random sources both with and without switching to assess its benefit---the \acro curve always tracks the {\em Top-Down} curve. This is the case for {\tt mawi} (Fig.~\ref{fig:switch:mawi}) and {\tt GAP-web} (Fig.~\ref{fig:switch:web}), where the {\em Top-Down} curve aligns exactly with the {\em Optimal} curve, confirming that the preprocessing decision is correct. When switching is enabled, \acro attempts to follow the {\em Optimal} curve by applying Eq.~\eqref{eq:switching} at each level. For most, \acro's decisions are near-optimal, e.g., on {\tt GAP-urand}~(Fig.~\ref{fig:switch:urand}), the curves are nearly indistinguishable, with the notable exception of {\tt GAP-kron} (Fig.~\ref{fig:switch:kron}), where a misclassification causes \acro to select the wrong mode at 2 out of 8 levels, resulting in a $1.69\times$ higher execution time than {\em Optimal}. On {\tt GAP-twitter} (Fig.~\ref{fig:switch:twitter}), the match is exact, demonstrating that $\eta = 10$ is an effective threshold on H100, although a graph-specific $\eta$ may extract the remaining performance headroom. We note, however, that $\eta = 10$ assumes that processing one VSS with TCs costs the same as processing 10 vertices with CUDA Cores, a ratio calibrated specifically for the Hopper GPU. This constitutes a limitation of \acro: the switching threshold is GPU-specific and may require retuning on other architectures.

Finally, we evaluate the optimizations of Section~\ref{sec:extension} for the multi-source BFS extension of \acro via the ablation study in Table~\ref{tab:msbfs_ablation}. (Naive) serves as the baseline, executing the SS-BFS kernel independently over multiple sources. Three further variants follow: (A) is Alg.~\ref{alg:msbfs_kernel} without reindexing or dynamic slice set tracking ($\mathbf{activeSets}$/$\mathbf{dirtySets}$); (AB) drops only the dynamic tracking; and (Full) is Alg.~\ref{alg:msbfs_kernel} in its entirety.

\begin{table}[h]
\centering
\caption{\small MS-BFS ablation study of \acro (in sec., total of 256 BFSs over random sources). (Naive): multiple SS-BFSs. (A): Alg.~\ref{alg:msbfs_kernel} without reindexing or dynamic slice set tracking. (AB): Alg.~\ref{alg:msbfs_kernel} without dynamic slice set tracking. (Full): Alg.~\ref{alg:msbfs_kernel}.}
\scalebox{0.75}{
\setlength{\tabcolsep}{2.6pt}
\begin{tabular}{c|l||r||rrr}
\multicolumn{2}{c}{}& \multicolumn{4}{c}{\textbf{Times (in seconds)}} \\ \cline{3-6}
& &
\multicolumn{1}{c||}{\textbf{\acro}} &
\multicolumn{3}{c}{\textbf{\acro}} \\
& \textbf{Graph} &
\multicolumn{1}{c||}{\textbf{(Naive)}} &
\multicolumn{1}{c}{\textbf{(A)}} &
\multicolumn{1}{c}{\textbf{(AB)}} &
\multicolumn{1}{c}{\textbf{(Full)}} \\
\midrule
\multirow{3}{*}{\rotatebox[origin=c]{90}{GAP}}
&{\tt{GAP-road}}      & 10.79 & 49.18  & 41.83  & 3.64 \\
&{\tt{GAP-twitter}}   & 1.73  & 1.80   & 1.47   & 1.11 \\
&{\tt{GAP-web}}       & 4.35  & 2.39   & 1.97   & 0.77 \\
\midrule
\multirow{7}{*}{%
  \rotatebox[origin=c]{90}{%
    \shortstack{%
      $|\mathcal{V}| \ge 23\mathrm{M}$\\
      $|\mathcal{E}| \le 2^{32}-1$
    }%
  }%
}&{\tt{nlpkkt240}}     & 2.30  & 10.61  & 10.12  & 1.68 \\
&{\tt{uk-2005}}       & 2.76  & 2.37   & 1.84   & 0.70 \\
&{\tt{it-2004}}       & 2.90  & 2.10   & 1.78   & 0.64 \\
&{\tt{europe\_osm}}   & 33.30 & 146.14 & 100.51 & 9.05 \\
&{\tt{com-Friendster}}& 3.00  & 4.87   & 4.20   & 2.75 \\
&{\tt{Spielman\_k600}}& 4.97  & 32.64  & 30.16  & 2.14 \\
&{\tt{webbase-2001}}  & 5.17  & 9.30   & 5.66   & 1.56 \\
\midrule\midrule
  & \multicolumn{1}{c||}{Min}     & 1.0$\times$ & 0.15$\times$ & 0.16$\times$ & 1.09$\times$ \\
  & \multicolumn{1}{c||}{Max}     & 1.0$\times$ & 1.82$\times$ & 2.21$\times$ & 5.65$\times$ \\
  & \multicolumn{1}{c||}{Geomean} & 1.0$\times$ & 0.52$\times$ & 0.65$\times$ & 2.69$\times$ \\
\end{tabular}}
\label{tab:msbfs_ablation}
\end{table}

\subsection{Overhead of \acro}
\label{subsec:overhead}

The preprocessing overhead of \acro is threefold: (1) constructing the CSC ({\em Compressed Sparse Columns}) of $\mathbf{A}$ from the edge list, which is an I/O-bound operation; (2) reordering the CSC matrix via either Alg.~\ref{alg:jaccard_with_windows} or RCM; and (3) constructing the BVSS data structure. The first overhead is shared by all state-of-the-art implementations listed in Table~\ref{tab:ssbfs_blest_vs_sota}, whereas the latter two are specific to \acro. Among the baselines, \cite{berrybees} also constructs a specialized data structure, BRS, whose construction cost is analogous to that of BVSS, as described in Sec.~\ref{sec:ds}. Table~\ref{tab:blest_overhead} reports the preprocessing costs for \acro for all the graphs. Whenever Alg.~\ref{alg:jaccard_with_windows} is selected, we set $W = 2^{16}$.

\begin{table}[h]
\centering
\caption{\small Preprocessing overhead of \acro (in sec.) across all graphs. The columns report: the ordering algorithm used, the cost of CSC construction from the edge list, reordering via \textsc{JaccardWithWindows} (J) or RCM (R), and BVSS construction.}
\scalebox{0.73}{
\setlength{\tabcolsep}{1.7pt}
\begin{tabular}{l|c|r|r|r||l|c|r|r|r}
\textbf{Graph} & 
\multicolumn{1}{c|}{\rotatebox{90}{\parbox{0.6cm}{\centering \textbf{Ord.}}}} & 
\textbf{CSC} & \textbf{Reord.} & \textbf{BVSS} &
\textbf{Graph} & 
\multicolumn{1}{c|}{\rotatebox{90}{\parbox{0.6cm}{\centering \textbf{Ord.}}}} & 
\textbf{CSC} & \textbf{Reord.} & \textbf{BVSS} \\
\midrule
{\tt{GAP-road}}     & R &    9.3 &    9.5 &   5.3 &
{\tt{GAP-twitter}}  & J &  534.6 &  388.1 &  72.1 \\

{\tt{GAP-web}}      & J &  340.8 &  871.8 &  20.5 &
{\tt{GAP-kron}}     & J & 1014.2 & 1804.4 & 159.6 \\

{\tt{GAP-urand}}    & J & 1294.0 & 1942.9 & 181.3 &
{\tt{nlpkkt240}}    & J &   79.9 &   28.9 &   8.4 \\

{\tt{uk-2005}}      & J &   85.8 &  306.3 &  10.4 &
{\tt{it-2004}}      & J &   97.4 &  199.6 &  12.9 \\

{\tt{europe\_osm}}  & R &   11.8 &   19.0 &  22.1 &
{\tt{com-Friend.}}  & J &  720.6 &  750.6 & 149.2 \\

{\tt{Spiel.\_k600}} & R &   42.9 &   23.5 &  15.7 &
{\tt{webb.-2001}}   & J &  100.6 &  429.1 &  15.1 \\

{\tt{kmer\_V1r}}    & R &   70.5 &  164.6 &  74.6 &
{\tt{mawi}}         & J &   62.0 & 1312.7 &  81.5 \\
\end{tabular}}
\label{tab:blest_overhead}
\end{table}

We finally evaluate the memory footprint of \acro in Table~\ref{tab:memory_footprint}. As evident from the table, the structures required for BVSS and the computational mechanics of \acro are remarkably compact, even for the largest real-world graphs. The size of level arrays in MS-BFS, however, grows proportionally with the number of concurrent BFS instances $\kappa$ and becomes the dominant memory consumer. For 4 graphs, this results in an out-of-memory (OOM) error on the NVIDIA H200, which is why these graphs are excluded from the MS-BFS experiments in Table~\ref{tab:msbfs_ablation}. We note that this limitation is inherent to all GPU-based MS-BFS frameworks, and reducing $\kappa$ straightforwardly can resolve this issue. 

\begin{table}[t]
\centering
\caption{\small Memory footprint of \acro (in GB) across all experimental graphs for the Multi-BFS workload ($\kappa = 256$ sources). The CSR column is only applicable to graphs for which direction switching is enabled; it is marked N/A otherwise.}
\scalebox{0.77}{
\setlength{\tabcolsep}{1.7pt}
\begin{tabular}{c|l||rrrrrrr}
& \textbf{} &
\multicolumn{1}{c}{\textbf{}} &
\multicolumn{1}{c}{\textbf{}} &
\multicolumn{1}{c}{\textbf{}} &
\multicolumn{1}{c}{\textbf{}} &
\multicolumn{1}{c}{\textbf{}} &
\multicolumn{1}{c}{\textbf{Active}} &
\multicolumn{1}{c}{\textbf{}} \\
& \textbf{Graph} &
\multicolumn{1}{c}{\textbf{CSR}} &
\multicolumn{1}{c}{\textbf{BVSS}} &
\multicolumn{1}{c}{\textbf{Frontier}} &
\multicolumn{1}{c}{\textbf{Queues}} &
\multicolumn{1}{c}{\textbf{Levels}} &
\multicolumn{1}{c}{\textbf{/Dirty}} &
\multicolumn{1}{c}{\textbf{Total}} \\
\midrule
\multirow{5}{*}{\rotatebox[origin=c]{90}{GAP}}
&{\tt{GAP-road}}      & N/A   & 1.77  & 2.30  & 0.02  & 24.52  & 0.10  & 28.71  \\
&{\tt{GAP-twitter}}   & 6.12  & 5.76  & 5.91  & 0.10  & 63.06  & 0.25  & 81.20  \\
&{\tt{GAP-web}}       & N/A   & 1.79  & 4.86  & 0.05  & 51.85  & 0.21  & 58.77  \\
&{\tt{GAP-kron}}      & 17.43 & 18.58 & 12.88 & 0.28  & 137.44 & 0.55  & 187.17 \\
&{\tt{GAP-urand}}     & 17.72 & 19.85 & 12.88 & 0.34  & 137.44 & 0.55  & 188.78 \\
\midrule
\multirow{9}{*}{\rotatebox[origin=c]{90}{$|\mathcal{V}| \ge 23M$, $|\mathcal{E}| \le 2^{32} - 1$}}
&{\tt{nlpkkt240}}     & N/A   & 1.28  & 2.69  & 0.03  & 28.67  & 0.12  & 32.77  \\
&{\tt{uk-2005}}       & N/A   & 0.88  & 3.79  & 0.04  & 40.41  & 0.16  & 45.28  \\
&{\tt{it-2004}}       & N/A   & 1.06  & 3.96  & 0.04  & 42.28  & 0.17  & 47.52  \\
&{\tt{europe\_osm}}   & N/A   & 3.77  & 4.89  & 0.05  & 52.13  & 0.21  & 61.05  \\
&{\tt{com-Friendster}}& 14.71 & 15.93 & 6.30  & 0.26  & 67.18  & 0.27  & 104.65 \\
&{\tt{Spielman\_k600}}& N/A   & 5.34  & 6.93  & 0.07  & 73.91  & 0.30  & 86.55  \\
&{\tt{webbase-2001}}  & N/A   & 1.48  & 11.34 & 0.09  & 120.98 & 0.49  & 134.38 \\
&{\tt{kmer\_V1r}}     & N/A   & 15.84 & 20.54 & 0.21  & 219.14 & 0.88  & 259.34 \\
&{\tt{mawi}}          & N/A   & 2.08  & 21.71 & 0.24  & 231.63 & 0.93  & 259.42 \\
\end{tabular}}
\label{tab:memory_footprint}
\end{table}

\section{Related Work}
\label{sec:related}

The idea of using linear-algebraic primitives in graph algorithms transformed graph analytics~\cite{combblas, GraphBLAS, graphblast}. This became prominent with the availability of MMA units in modern accelerators. BerryBees~\cite{berrybees} is an early effort to realize BFS using TCs via a bitmap-based data structure. In contrast to \acro, it uses both SpMV and SpMSpV as a primitive and switches from the latter to the former whenever the frontier is sufficiently dense\footnote{This mechanism does not switch between push and pull. Instead, it improves the work efficiency by skipping pulls for vertices whose incoming neighbours contain no frontier vertices.}. Both variants of their implementation, however, suffer from major bottlenecks as explained in this work.
BerryBees' SpMV applies a \emph{frontier-oblivious slice-set distribution}, which in turn severely exacerbates load imbalance across the warps. When the frontier becomes sparse, BerryBees switches to SpMSpV. 

BerryBees tries to solve the load-balancing problem in the SpMSpV by traversing the frontier and redistributing the slice sets based on the number of slices. This per-level, frontier-dependent redistribution limits the benefits of SpMSpV and highlights the importance of the proposed BVSS data structure, which achieves near-perfect load balance \emph{by construction}, without requiring the algorithmic layer to be aware of or to react to such dynamic load variations.

Gunrock~\cite{gunrock} is a GPU library designed to balance performance, scalability, and programmability by expressing the graph algorithms through a data-centric ``frontier” abstraction and a bulk-synchronous loop. It applies GPU-specific optimizations such as load-balanced workload mapping, idempotence to reduce atomics under concurrent discoveries, and push vs. pull direction switching. Hence, it specializes the execution flow with respect to the graph at hand. The literature, and also this work, follows the message that there is no single “best” GPU BFS frontier expansion policy that does not vary w.r.t. the graph structure. GSWITCH~\cite{gswitch} dynamically reconfigures (with low overhead) the execution by selecting among optimization patterns, most critically push/pull switching, plus complementary choices such as load-balancing and active-set representation. It uses runtime/workload features to decide when to switch and which kernels to run, searching a BFS variant space in each iteration.

\section{Conclusion and Future Work}
\label{sec:conclusion}

In this paper, we presented \acro, an ultra-efficient framework for executing BFS, Multi-Source BFS, and Closeness Centrality on modern GPUs at a speed unprecedented in the literature. By systematically engineering every stage of the pipeline and developing novel approaches tailored specifically to the unique characteristics of BFS, we establish \acro as the fastest BFS implementation in the era of Tensor Cores, achieving a $5.9\times$ geomean speedup over the prior state-of-the-art across 14 diverse real-world graphs. Although \acro demonstrates strong performance across the vast majority of experimental configurations, the direction-switching mechanism occasionally misclassifies the optimal execution mode on certain graphs, leaving extractable performance unrealized. Developing a more robust, potentially graph-adaptive switching criterion that minimizes misclassification without incurring prohibitive preprocessing cost constitutes a natural direction for future work.

\section*{Acknowledgments}

The numerical calculations reported in this paper were fully performed at 1) TUBITAK ULAKBIM, High Performance and Grid Computing Center (TRUBA resources) and 2) the EuroHPC Joint Undertaking (EuroHPC JU) supercomputer MareNostrum 5, hosted by the Barcelona Supercomputing Center (BSC). Access to MareNostrum 5 was provided through a national access call coordinated by the Scientific and Technological Research Council of Turkey (TÜBİTAK). We are grateful to TÜBİTAK and EuroHPC JU for providing access to these resources and supporting this research.

\bibliographystyle{IEEEtran}
\bibliography{main.bib}

\begin{IEEEbiography}
[{\includegraphics[width=1in,height=1.25in,clip,keepaspectratio]{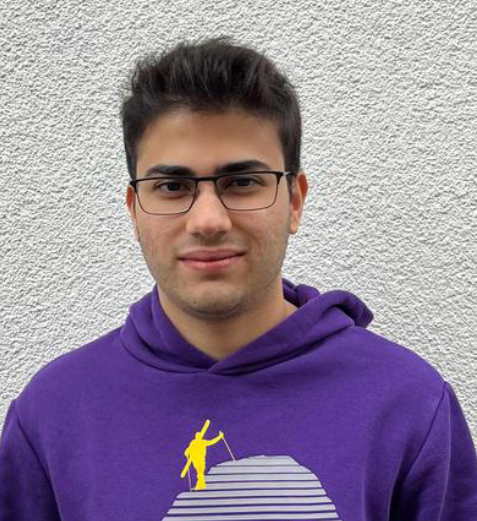}}]{Deniz Elbek} is an undergraduate student in Computer Science and Engineering at Sabanci University. His areas of research focus on High Performance Computing, Parallel Algorithms, and Massively Parallel Computer Architecture.
\end{IEEEbiography}

\begin{IEEEbiography}[{\includegraphics[width=1in,height=1.25in,clip,keepaspectratio]{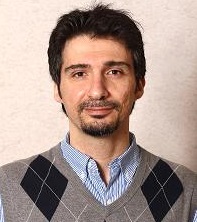}}]{Kamer Kaya} is an Associate Professor of Computer Science at Sabanci University. His research interests are Parallel Algorithms, High Performance Computing, and Graph and Sparse Matrix Algorithms.
\end{IEEEbiography}

\end{document}